\documentclass[journal]{IEEEtran}
\ifCLASSINFOpdf
   \usepackage[pdftex]{graphicx}
   \graphicspath{{../pdf/}{../jpeg/}}
   \DeclareGraphicsExtensions{.pdf,.jpeg,.png}
\else
   \usepackage[dvips]{graphicx}
   \graphicspath{{../eps/}}
\fi
\usepackage{array}
\usepackage{multirow}

\usepackage{amsmath}
\usepackage{booktabs}
\usepackage[caption=false,font=footnotesize]{subfig}
\usepackage[noadjust]{cite}
 
\usepackage{textcomp}
\usepackage[flushleft]{threeparttable}
\usepackage[dvipsnames]{xcolor}
\begin{document}
%
% paper title
% Titles are generally capitalized except for words such as a, an, and, as,
% at, but, by, for, in, nor, of, on, or, the, to and up, which are usually
% not capitalized unless they are the first or last word of the title.
% Linebreaks \\ can be used within to get better formatting as desired.
% Do not put math or special symbols in the title.
\title{A Review of 5G Front-End Systems Package Integration}
%
%
% author names and IEEE memberships
% note positions of commas and nonbreaking spaces ( ~ ) LaTeX will not break
% a structure at a ~ so this keeps an author's name from being broken across
% two lines.
% use \thanks{} to gain access to the first footnote area
% a separate \thanks must be used for each paragraph as LaTeX2e's \thanks
% was not built to handle multiple paragraphs
%

\author{Atom O. Watanabe,~\IEEEmembership{Student Member,~IEEE,}
		Muhammad Ali,~\IEEEmembership{Student Member,~IEEE,} \\
        Sk Yeahia Been Sayeed,
        Rao R. Tummala,~\IEEEmembership{Life Fellow,~IEEE,}
        P. Markondeya Raj,~\IEEEmembership{Senior Member,~IEEE.}
        \\        % <-this % stops a space
\thanks{A. O. Watanabe, M. Ali, and R. Tummala are with the Department
of Electrical and Computer Engineering, Georgia Institute of Technology, Atlanta,
GA, 30332 USA e-mail: atom@gatech.edu.}% <-this % stops a space
\thanks{S. Y. B. Sayeed and P. M. Raj are with the Department of Biomedical Engineering, Florida International University, Miami, FL 33174 USA.}% <-this % stops a space
%\thanks{Manuscript received April 19, 2005; revised August 26, 2015.}
}

% note the % following the last \IEEEmembership and also \thanks - 
% these prevent an unwanted space from occurring between the last author name
% and the end of the author line. i.e., if you had this:
% 
% \author{....lastname \thanks{...} \thanks{...} }
%                     ^------------^------------^----Do not want these spaces!
%
% a space would be appended to the last name and could cause every name on that
% line to be shifted left slightly. This is one of those "LaTeX things". For
% instance, "\textbf{A} \textbf{B}" will typeset as "A B" not "AB". To get
% "AB" then you have to do: "\textbf{A}\textbf{B}"
% \thanks is no different in this regard, so shield the last } of each \thanks
% that ends a line with a % and do not let a space in before the next \thanks.
% Spaces after \IEEEmembership other than the last one are OK (and needed) as
% you are supposed to have spaces between the names. For what it is worth,
% this is a minor point as most people would not even notice if the said evil
% space somehow managed to creep in.

% The paper headers
\markboth{Preprint September~2020}%
{Shell \MakeLowercase{\textit{et al.}}: Bare Demo of IEEEtran.cls for IEEE Journals}
% The only time the second header will appear is for the odd numbered pages
% after the title page when using the twoside option.
% 
% *** Note that you probably will NOT want to include the author's ***
% *** name in the headers of peer review papers.                   ***
% You can use \ifCLASSOPTIONpeerreview for conditional compilation here if
% you desire.

% If you want to put a publisher's ID mark on the page you can do it like
% this:
%\IEEEpubid{0000--0000/00\$00.00~\copyright~2015 IEEE}
% Remember, if you use this you must call \IEEEpubidadjcol in the second
% column for its text to clear the IEEEpubid mark.

% use for special paper notices
%\IEEEspecialpapernotice{(Invited Paper)}

% make the title area
\maketitle

% As a general rule, do not put math, special symbols or citations
% in the abstract or keywords.
\begin{abstract}
Increasing data rates, spectrum efficiency and energy efficiency have been driving major advances in the design and hardware integration of RF communication networks. In order to meet the data rate and efficiency metrics, 5G networks have emerged as a follow-on to 4G, and projected to have 100X higher wireless date rates and 100X lower latency than those with current 4G networks. Major challenges arise in the packaging of radio-frequency front-end modules because of the stringent low signal-loss requirements in the millimeter-wave frequency bands, and precision-impedance designs with smaller footprints and thickness. Heterogeneous integration in 3D ultra-thin packages with higher component densities and performance than with the existing 2D packages is needed to realize such 5G systems. This paper reviews the key building blocks of 5G systems and the underlying advances in packaging technologies to realize them. 
\end{abstract}

% Note that keywords are not normally used for peerreview papers.

% For peer review papers, you can put extra information on the cover
% page as needed:
% \ifCLASSOPTIONpeerreview
% \begin{center} \bfseries EDICS Category: 3-BBND \end{center}
% \fi
%
% For peerreview papers, this IEEEtran command inserts a page break and
% creates the second title. It will be ignored for other modes.
\IEEEpeerreviewmaketitle

\section{Path to 5G}
\label{sec:intro}

\IEEEPARstart{M}{illimeter}-wave (mm-wave) telecommunication for fifth-generation (5G) communication is finally becoming a reality. 5G will impact our lives more dramatically than any technology shift since the internet itself simply because 5G leads us to a fully connected world. The realization of 5G drives new packaging technologies compared to  conventional RF packaging. This review paper thoroughly covers key research the advances and their technology integration towards the 5G mm-wave packaging. It begins with a brief introduction of 5G systems, applications, and performance attributes for demystifying the need for advanced packaging building blocks (Section \ref{sec:intro}). Section \ref{sec:architecture} briefly covers the system architecture perspective of 5G technologies. Section \ref{sec:challenges} highlights package-level technical challenges, followed by the key technology elements for heterogeneous package integration as the key enabler for 5G (Section \ref{sec:keytech}). Section \ref{sec:demo} covers the packaging trends in RF front-end, material development for 5G, integration of passive components, and antenna-integrated packages. This review paper also discusses recent demonstrations of 5G mm-wave packages, followed by the introduction of the potential of sixth-generation (6G) communications and associated opportunities in packaging (Section \ref{sec:6G}).

\subsection{5G vs 4G Communications}
\label{subsec:4G5G}
5G is anticipated to be disruptive and life changing as it promises a significantly broader range of applications than its previous counterparts: 4G and long term evolution (LTE), and LTE Advanced (LTE-A/4.5G) networks. Voice was the sole purpose of earlier cellular networks such as 2G and Enhanced Data Rates for GSM Evolution (EDGE). Design changes subsequently were needed because the high latency of 2G and EDGE systems was unable to fulfill the need for mobile internet access. This resulted in the dawn of 3G network, which not only provided more capacity for voice but also introduced fast data services as the network was packaged-switched unlike its circuit switched predecessors. 4G was evolutionary to 3G as it increased the capacity by providing higher data rates and added more services at an affordable cost to the consumer \cite{Qualcomm_EvolutionofMobileTech}.

Unlike LTE which is an evolutionary measure, 5G is an entire new network with a broad range of applications and use-cases \cite{Introducing5GTech_Thales}. That is why new radio (NR) is the term coined for 5G as it uses denser modulation schemes, waveforms and pairing with current technologies to meet the ever-growing demands of communication networks. The transition to 5G from LTE is already much faster than it was for 3G to 4G as technologies have rapidly advanced. The users of 5G go beyond consumer-centric networks and include businesses, services, utilities, cities, and beyond. Speed, capacity, number of connected devices, latency and reliability are the key metrics that are driving the advent of 5G. Given these diverse requirements, the challenge is that 5G network is expected to support them in a flexible way with high fidelity \cite{5GUseCasesNokia}. Major improvements in 5G compared to current LTE-A networks are listed in Table \ref{tab:compare_5G_LTE_A}.

\begin{figure}[!t]
	\centering
	\includegraphics[width=0.48\textwidth]{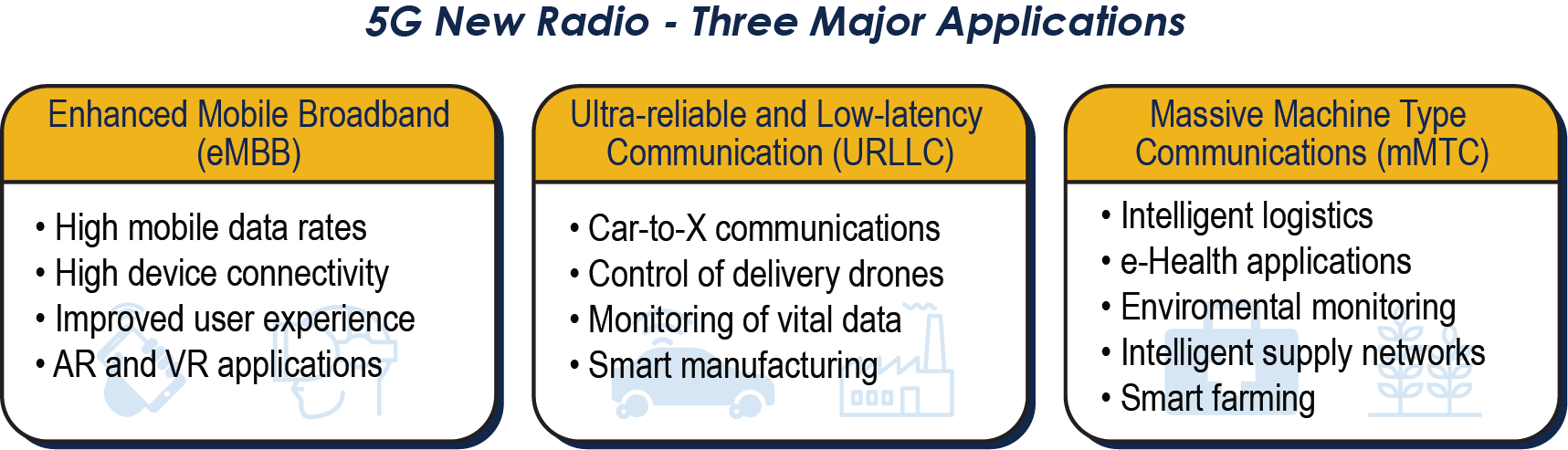}
    \caption{Major classes of applications with 5G networks.}
	\centering
	\label{fig:eMBB, URLLC, mMTC}
\end{figure}

{
\newcolumntype{a}{>{\centering\arraybackslash} m{2.0cm} }
\newcolumntype{e}{>{\centering\arraybackslash} m{2.8cm} }
\begin{table}[b]
\centering
\renewcommand{\arraystretch}{1.5}
\caption{Improvements in 5G Technology compared to LTE-A}
\label{tab:compare_5G_LTE_A}
\begin{tabular}{aee}
\hline
\textbf{Parameter}       & \textbf{5G}       & \textbf{Improvement compared to LTE-A} \\ \hline\hline
Device/km$^2$      & 10\textsuperscript{6} \par devices/km\textsuperscript{2}    & 10-100X                       \\ \hline
Latency         & \textless{}1 ms    & 10X                           \\ \hline
Data Rate       & 10-20 Gbps         & 10-20X                        \\ \hline
Frequency Range & 600 MHz to mm-Wave & 600 MHz to 5.925 GHz          \\ \hline
Channel Bandwidth & \begin{tabular}[e]{@{}e@{}}100 MHz for $<$6 GHz\\ 400 MHz for mm-wave\end{tabular} & 5X-20X \\ \hline
\end{tabular}
\end{table}
}

\subsection{5G Applications (eMBB, URLLC, mMTC)}
\label{sec:5G_applications}
Enhanced mobile broadband (eMBB), ultra-reliable low latency communication (URLLC) and massive machine type communication (mMTC) are the three use cases of 5G air interface as shown in Fig. \ref{fig:eMBB, URLLC, mMTC}. The focus of eMBB is to support ever-increasing demand of end user data rates by increasing system capacity \cite{Explore5GNR_UseCase_Qualcomm}. This is achieved by pushing the envelope of current limitations of the frequency spectrum, the primary being available bandwidth. This limitation is overcome by licensing available bandwidth in sub-10 GHz range as well as opening mm-wave frequencies, where wide bandwidth can be easily allocated. Antenna arrays with multiple antenna elements enable massive multiple input multiple output (MIMO) and beamforming, which multiplies the capacity of a network by exploiting multi-path propagation. eMBB is the initial phase and face of 5G deployment and it is helping to develop modern broadband use-cases such as Ultra-HD and 360$^\circ$ streaming, and emerging augmented reality (AR) and virtual reality (VR) media and applications.

The ability to process and harmonize various inputs for fast response falls under the domain of URLLC. It provides ultra response connections with $<$1-ms latency and 99.9999\% availability of connection along with support of high-speed mobility for mission critical applications \cite{UltraRel_Qualcomm}. Some of the use-cases of URLLC services include security, vehicle-to-vehicle (V2V) and vehicle-to-everything (V2X) applications, healthcare, utilities, cloud and real-time monitoring situations. The primary advantage of URLLC is not speed but its unfettered reliability to support robust and autonomous real-time decision making \cite{bennis2018ultrareliable}.

Finally, mMTC service targets robust and cost-sensitive connection of billions of devices such as Internet of Things (IoT) with long-time availability and low power consumption. As the name suggests, it is primarily for machine-to-machine (M2M) applications with minimal human interaction. This service considers IoT with low data rate for a large number of connected devices such as sensors with long range and low maintenance times \cite{5G_RF_for_dummies_Qorvo}. Essentially, mMTC entails a large mesh of low-cost, densely connected devices.

\subsection{5G Frequency Bands}
\label{subsec:freqbands}
The 3rd Generation Partnership Project (3GPP) unites seven telecommunications standard development organizational partners in six countries and provides their members with a stable environment to produce the reports and specifications that cover cellular telecommunications technologies. 3GPP has defined two frequency range designations: Frequency Range 1 (FR1) and Frequency Range 2 (FR2). FR1 covers low- and mid-frequency bands whereas FR2 is solely for mm-wave bands. The frequency range of FR1 and FR2 are defined below:

\begin{itemize}
    \item FR1: 450 MHz - 6000 MHz (3GPP Rel. 15 V15.4.0 \cite{3gpp.38.104_V15.4_2018}), 410 MHz - 7125 MHz (3GPP Rel. 15 V15.5.0 \cite{3gpp.38.104_V15.5_2019}) 
    \item FR2: 24250 MHz - 52600 MHz
\end{itemize}

Till 3GPP Release 15 V15.4.0 \cite{3gpp.38.104_V15.4_2018}, the maximum frequency range of FR1 was 6000 MHz hence the name ``sub-6 GHz" became a norm but it was changed to 7125 MHz in 3GPP Release 15 v15.5.0. FR2 has remained the same till the latest report (3GPP Release 16 v16.3.0) \cite{3gpp.38.104_V16.3_2020}. . The 5G spectrum is classified into low-, mid- and high-bands. They are defined below:

\begin{itemize}
    \item Low-band: $<$1 GHz
    \item Mid-band: sub-7 GHz (1 GHz - 7.125 GHz)
    \item High-band: mm-wave (above 24 GHz)
\end{itemize}

In the US, Federal Communications Commission (FCC) regulates the communications by radio, television, wire, satellite, and cable. FCC has its own categorization of 5G spectrum for low-, mid- and high-bands licensed in the US \cite{FCC_5G_bands}.

With the rapid need of 5G roll-out, two approaches are given by 3GPP: non-standalone (NSA) and standalone (SA). NSA is the early version of 5G NR as it uses LTE radio access network (RAN) and core with additional support for 5G low- and mid-bands (sub-7 GHz). This variant is for fast-to-launch approach favored by many carriers throughout the world. On the other hand, the long-term variant, SA has advantages in terms of simplicity and improved efficiency of 5G next gen core, lower costs, steady improvement of performance in the entire network while enabling URLLC and mMTC use-cases. To be precise, SA gives 5G NR the ability for independent deployment as it is an end-to-end solution with solid basis to unleash its full potential. However, it has multiple time-consuming and cost-demanding challenges associated with it such as building new 5G infrastructure \cite{Making5G5Commercial_Qualcomm}.

As an inference from the two methods to approach 5G, it can be asserted that NSA 5G requires more incremental improvements over its SA counterpart. 4G LTE had 40 bands and the 5G cellular networks support even more low- and mid-bands, resulting in complexity in hardware design. The first phase of 5G deployment and its implementation has introduced challenges by adding more bands in FR1. Similarly, FR2 requires addition of new hardware, resulting in added complexity and miniaturization challenges. This has driven advances in all aspects of the system hardware: from ICs to devices and from cellular devices to base stations. The low-band 5G using 600 MHz spectrum (band n71) was launched in the US in December 2019 \cite{T-Mobile_5G_Launch, Samsung_GalaxyS105G, Samsung_GalaxyS205G}.

\section{5G System Drivers} 
\label{sec:architecture}
{   \newcolumntype{a}{>{\centering\arraybackslash} m{2.6cm} }
    \newcolumntype{q}{>{\centering\arraybackslash} m{5.4cm} }
    \begin{table*}[t]
	\centering
	\caption{Technology metrics for 5G communication systems*}
	
	\begin{tabular}{aaqq}
    \hline 
                                            & Downlink \par (Base station)                                                    & Uplink (CPE)                                                                                                                                                              & User equipment (UE)                                                                                                                                                                                                                      \\ \hline \hline
Antenna and \par module size & 70$\times$70$\times$2.7 mm$^3$                         & 450--1400 mm$^2$   \par Substrate thickness: 1.5 mm                       & 20$\times$5$\times$2   mm$^3$  (Qualcomm, QTM052)                                                                                                                     \\ \hline
Antennas                                    & 64 -- 256                                                                    & 16 -- 32                                                                                                                                                                  & 4 -- 8                                                                                                                                                                                                                                   \\ \hline
PA power                                    & 33 dBm                                         & 19 dBm                                                                                                                                      & 10-15 dBm (6-8 dBm usually)   \par 28 nm CMOS   \par DC power for four elements: 360 -- 380 mW   \par Power consumption per channel $<$ 100 mW \\ \hline
Antenna gain                                & 27 dBi EIRP\textsuperscript{1} of    50 dBm (IBM) & 18--21 dBi for 8$\times$4   array  patch antenna  with grounded rings & 20 dBi for 2$\times$2 antenna array\\ \hline
End-to-End loss &  -- & --  & 2.5 dB                                                                                                                                                                                                                                   \\ \hline
Pathloss  & 135 dB & 135 dB & -- \\ \hline
Received power & -75 dBm & -90 dBm & -140 dBm\textsuperscript{2} \\ \hline
SNR per RX Element & 5 dB & -15 dB & 6.2 dB\textsuperscript{3} \\ \hline
Rx gain  & 21 dBi  & 27 dBi & $\sim$10 dBi  \\ \hline
Rx SNR after gain  & 26 dB  & 12 dB   & --\\ \hline
    \end{tabular}
	 \begin{tablenotes}
	\item[1] \footnotesize *Parts of this table are taken from \cite{2020Cameron}
	\item[2] \footnotesize \textsuperscript{1}Effective isotropic radiated power (EIRP)
	\item[3] \footnotesize \textsuperscript{2}Minimum reference signal received power (RSRP)
	\item[4] \footnotesize \textsuperscript{3}Minimum performance requirements for 16-QAM modulation format and 0.48 code rate for Rank 1 (beamforming), defined by 3GPP \cite{3gpp.38.101-4_V16.0.0_2020}
	\end{tablenotes}
	\label{tab:architecture}
	\end{table*}
	}
	
Systems that operate in 5G are classified as User Equipment (UE), Customer Premise Equipment (CPE), and Base Station or infrastructure. They have varying needs, size, and power constraints. UE is driven by miniaturization and reduced power while infrastructure equipment is geared towards high gain, communication range, massive-MIMO needs, and broadband. The technology metrics for these classes are compared in Table \ref{tab:architecture}.

Complexity of 5G systems is a direct result of the addition of features to established hardware, which requires tighter integration, inter-operability, and shielding of all components and devices in a module. Similarly, signals at higher frequencies experience higher losses in free-space due to higher attenuation as predicted by Friis transmission equation. A solution to mitigate this attenuation is to use highly-focused, narrow beams (beamforming and MIMO) to communicate with the target nodes in an efficient manner. This is accomplished by using higher order antenna arrays which are driven by multiple active components. MIMO beamforming technique is primarily classified into three categories: digital, analog, and hybrid. In analog beamforming, amplitude and phase variation is controlled as modulated signals at transmit end, while the received signals from each antenna are summed before the analog-to-digital (ADC) conversion. In digital beamforming, the amplitude/phase variation is controlled as demodulated signals at transmit end while received analog signals are processed after the ADC. Hybrid beamforming technique takes the advantages of both analog and digital beamforming types, which include high flexibility, channel configuration, and low cost and power consumption. The base station can thus serve multiple users in a timeslot.

Analog beamforming is preferred in mobile systems because a single RF chain can address the antenna elements while beam control is performed with passive elements. Digital beamforming consumes higher power with complex hardware but can have full beam control with several simultaneous beams and multiple apertures for future ultra-wide-band communication and software-defined radio features for smart and dynamic spectrum hopping and band-selection. In such innovative digital beamforming approach, the antenna outputs are encoded and multiplexed, and digitized with a single ADC, while demultiplexing and beamforming are performed in the digital domain \cite{2017Venkatakrishnan1}. Novel onsite coding approach that combines frequency and code multiplexing is developed to reduce the power requirements of digitizer by 10 to 32X \cite{2017Venkatakrishnan2}. In order to attain in-band duplex operation to effectively utilize the current-congested spectrum allocation, systems employ time (TDD) or frequency division duplexing (FDD) where the interference between transmit and receive signals is suppressed by differentiating them in time or frequency. However, both require doubling the time or frequency resources. This limitation is addressed with simultaneous transmit and receive (STAR) systems to better utilize the 5G spectrum. In-band full-duplex operation is dependent upon the cancellation of interference between the strong Tx and Rx using techniques such as cross-polarization array elements, and filtering in the analog processing to reduce Tx harmonics, and digital signal processing to filter out multipath domains \cite{2018Venkatakrishnan2, 2015Scherer}.

The need for low losses demands close integration between antennas with high-power PA and low-noise-figure LNA. As the antenna element size and pitch are comparable to the wavelength, antenna-in-package solutions make this approach feasible with mm-wave communications. Integrating antenna into a packaging substrate reduces the overall interconnect length between RFIC and antennas, and, therefore, mitigates the feed-line loss and enhances the antenna efficiency \cite{2019Gu}. In conjunction with the prior popularity of AiP in the 60-GHz band designed for radar, the trend of AiP for 28-GHz or 39-GHz has been more prominent for recent years for consumer electronics. 

The choice and integration of TRx technologies vary with the number of antenna elements and antenna gain to attain the required signal power. The radio-frequency front-end (RFFE), control and calibration circuits are integrated into the CMOS \cite{2018Kim}, SiGe BiCMOS \cite{2017Sadhu,2017Kibaroglu} or GaAs \cite{2016Curtis} transceiver dies, and are connected to the antennas, passive components, and power circuitries using routing layers in the package, microbumps, and through package vias (TPVs). Lower number of elements utilize GaN while higher antenna elements are driven by SiGe BiCMOS, and much higher antenna elements with bulk CMOS. Highly-integrated wireless transceivers further benefit from deep sub-micron to 28 nm technology to achieve single-chip CMOS solutions.

CMOS and SiGe PAs are optimized for 25 dBm power and 30\% power efficiency while GaAs achieves 30-35 dBm and GaN providing capabilities beyond 45 dBm. A 64-element array needs up to 5 W per PA to reach 240 W. Each PA must operate in the linear regime to reduce the error vector magnitude, which is more challenging for 64-array antennas. As the efficiency of PA modules with the required 8-dB back-off is estimated as 40\%, the increased power dissipation from the combined analog and digital processing requires integrated thermal management.

\section{Packaging Challenges in mm-wave 5G}
\label{sec:challenges}

Packaging of 5G systems needs integration of RF, analog and digital functions along with passives and other system components in a single module. These systems ideally exemplify the heterogeneous integration trend. This becomes more important for 5G because of several reasons: a) integration of antennas with transceiver ICS and associated passive and RF power divider netweorks, b) addition of sub-6 GHz (FR1) in the short-term with advances in packaging technologies, c) new mm-wave bands (FR2) drive the integration of new filters and diplexers along with broadband power amplifiers and switches, d) the add-on modules to the existing RFFE put additional emphasis on miniaturization and component integration. Proximity of the transceiver and front-end module is also important to reduce the size and losses. This is achieved by integration of antennas with the RF module as well as simultaneous modeling of a heat dissipation solution to keep active components in acceptable thermal conditions. Integration of power amplifiers with antenna arrays needs to address the challenges with size, cost, and performance \cite{2019Westberg}. These challenges translate to multi-layer fabrication with fine-line features and precise layer-to-layer registration, advanced low-loss materials to reduce conductive losses and co-simulation of circuit, device, package, and thermal solutions. The emerging 3D package integration solutions also underscore the need for isolation between the various circuit blocks. Because of the deployment of such high-power amplifiers and a large antenna array in millions of base-stations, cost needs to be addressed for high-volume manufacturing.

\subsection{High-density integration of mm-wave components for 5G RF front-end modules}
\begin{figure}[t]
		\centering
		\includegraphics[width=0.48\textwidth]{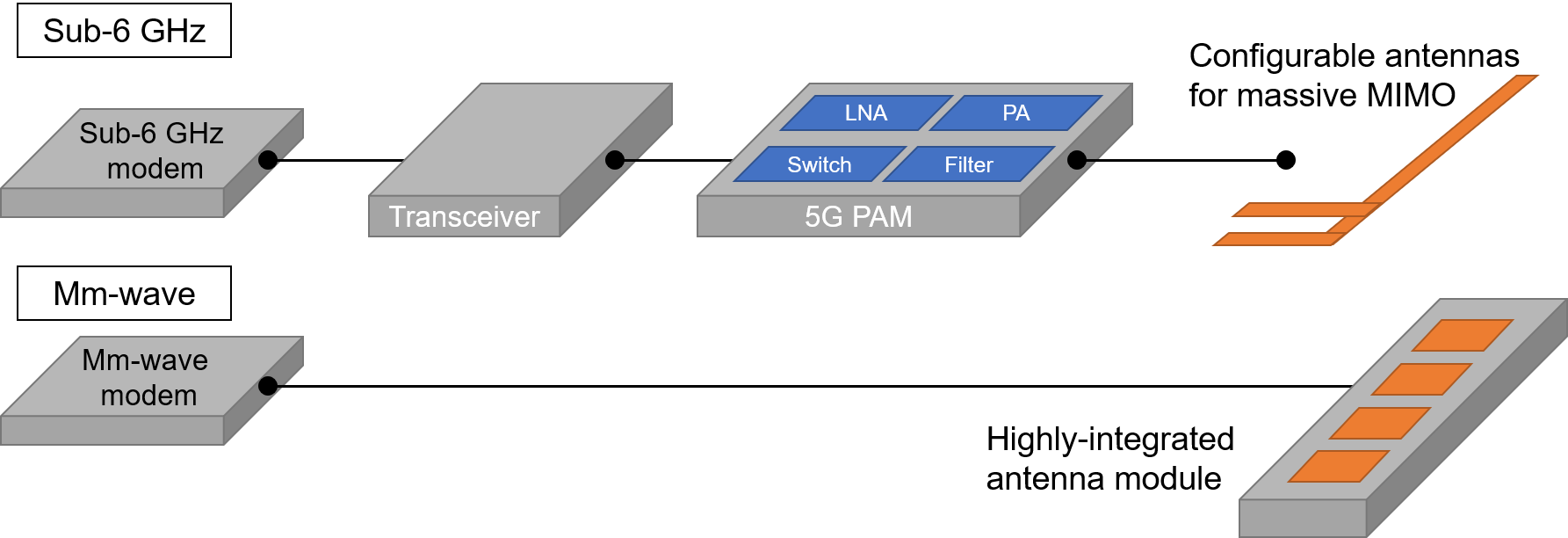}
    	\caption{An example of tight integration in RFFE architectures for 5G electronics.}
		\centering
		\label{fig:architecture}
\end{figure}

5G mm-wave modules entail tight integration of antenna arrays, transceiver ICs, power-management ICs, the stack of logic-memory, and surface mounted passive components, as depicted in Fig. \ref{fig:architecture}. The reported thickness of mm-wave packaging substrates varies from 0.15 mm to 1.2 mm. The variation results mainly from the antenna-in-package requirements. Packages without antenna could employ thin substrates below 300 $\mu$m. However, antenna-integrated modules require higher thickness as antenna arrays offer higher bandwidth with a thicker substrate due to more separation of ground plane from antenna patches. The thickness of entire modules (packaging-substrate thickness $+$ mold height) also varies from 0.5 mm to 2 mm. In addition to the thickness requirement for antenna, the system components and interconnects (Fig. \ref{fig:architecture}) determine the total thickness and area of the entire packages.

\subsection{Novel circuit design techniques}
Integration of multiple components into a package require more complicated signal routing and power delivery network in a limited space, to attain the required signal and power integrity and the control of electromagnetic-interference (EMI).  The dielectric thickness for signal routing and power distribution can be 15 $\mu$m and lower \cite{2018Watanabe}. 

\subsection{Multi-physics analysis and modeling}
Ideally, devices, component, and interconnects should maintain stable performance under various signal, power, and thermomechanical loads. As discussed in \cite{2019Geyik,2019Sivapurapu}, the packaging structure affects not only the signal and power distribution but also the electrical performance of each component or device. This demands accurate and reliable multi-physics modeling and analyses from the electrical, materials, chemical and mechanical standpoints.

\subsection{Co-Simulation of package, circuit and device}
5G mm-wave technology, as discussed in Section \ref{sec:architecture}, entails co-design of phased antenna array with transceiver or beamforming ICs to meet the power distribution and signal integrity requirements, and passive components to support all the functions with minimal parasitics and interference. The co-design should also consider electromagnetic compatibility and thermal design, as exemplified by Gu \textit{et al.} \cite{2019Gu}. 

\subsection{Novel Materials with good electrical and thermal properties}
In conjunction with the traditional material requirements for dielectrics in IC packaging such as compatible CTE and Young’s modulus, the new class of requirements especially for AiP leads to the development of low-loss dielectric materials. Low-loss materials are translated to materials with low dissipation factor, or loss tangent (tan $\delta$). Low tan $\delta$ mitigates dielectric loss in in-package interconnects, feedlines, and antennas, and thus increases the antenna efficiency. The high losses in the power amplifiers make thermal management more challenging. The recent notable material development is discussed in Section \ref{subsec:material}. In addition, there are studies on advanced RF conductors other than conventional copper, such as graphene \cite{2017Wang}, graphene composite \cite{2018Cheng}, copper paste \cite{2020Wang}, artificial magnetic conductors \cite{2017Lin}, metasurface \cite{2019Lee}, and metaconductors \cite{2018Hwangbo}. These materials provide lower resistivity, superior impedance control, or lower manufacturing cost over the traditional copper patterning. 

\subsection{Process challenges for high precision patterning}
Requirements for RFFE packaging differ from those for high-performance computing. RFFE conventionally prioritizes impedance control more than the high number of I/O or the bandwidth of the interconnects. However, emerging heterogeneous-integration trend to support massive MIMO will require finer-pitch I/O between the chip, the package, or antenna array. It is, therefore, imperative to manage both impedance control and minimize tolerances of the manufacturing processes for small line width and spacing (L/S) down to 5/5 $\mu$m to 2/2 $\mu$m by 2025 and 1/1 $\mu$m by 2030 \cite{2017Yole}. The emergence of AiP in front-end packages underscores the importance of multi-layer fabrication and layer-to-layer alignment with the precisely-controlled dielectric thickness. The fabrication accuracy of multi-layers is one of the key metrics to obtain good model-to-hardware correlation. Furthermore, metallization quality plays an important role. The surface roughness of the conductor degrades the signal quality, which needs to be incorporated in the process design to mitigate the system interconnect loss in conjunction with the effort to lower the dielectric loss.   

\subsection{Thermal management for 5G}
Reduced distance and the increased number of component density call for integrated heat-spreading structures from the thermal-management standpoint. 5G mm-wave technology is expected to consume more power than the previous generations of wireless technologies \cite{2019Kim}. Efforts to address thermal-management challenges have been made both by the academia  \cite{2018Chen,2019Yook} and industry \cite{2019Hwang,2019Gu,2018You} mostly with electrical and thermal co-design and analysis.

\section{5G Packaging Technology Building Blocks}
\label{sec:keytech}

\subsection{Packaging Trends for 5G Systems}
\label{subsec:trend}
\begin{figure}[t]
		\centering
		\includegraphics[width=0.3\textwidth]{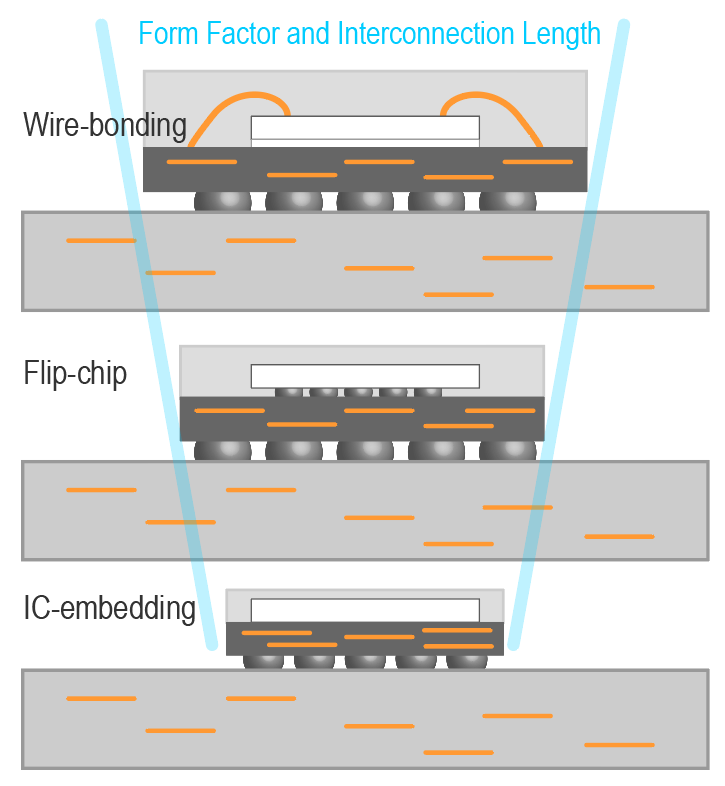}
    	\caption{Trends of interconnection assembly methods in RF/mm-wave technology for miniaturization. Wire-bonding as the most traditional low-I/O interconnection, flip-chip interconnection down to 25 $\mu$m, and IC-embedding or fan-out interconnection.}
		\centering
		\label{fig:interconn}
	\end{figure}

System-level packaging in the mm-wave technology are partitioned into baseband modules and antenna-integrated transceiver modules. In such packages, interconnections between ICs and other elements such as antenna, passive components, and PCBs must satisfy several requirements. One of the most critical requirements is the impedance control especially in the analog domain. In the mm-wave antenna-in-package solutions, interconnections between transceiver ICs and antennas should result in low insertion loss and acceptable return loss over the frequency range of interest. The other key requirement is the form factor; two types of interconnect techniques are widely available in the packaging industry, and a third technique has been rapidly emerging during the past decade. The two conventional techniques are wire-bonding (Fig. \ref{fig:interconn}a) and flip-chip interconnections (Fig. \ref{fig:interconn}b), whereas the emerging technique is referred to as IC-embedding or fan-out packaging (Fig. \ref{fig:interconn}c). Although flip-chip and fan-out interconnections have originally been developed for high-performance computing (HPC) or mobile processor applications, their interconnection attributes such as fine pitch and low electrical parasitics are getting critical in RF/mm-wave packages such as baseband modules and antenna-integrated modules.

Despite the maturity and cost effectiveness in the packaging industry, interconnection using the wire-bonding technique has been identified as one of the key challenges in system-on-package (SoP) because of the significant signal loss and impedance discontinuity caused by the bond wire, which degrades the performance of the RF/mm-wave system chain \cite{2009Zhang}. Interconnection using the flip-chip technique offers better performance than wire-bonding as the bump height is smaller than the length of bond wires. The flip-chip technique is also preferable for small form factor, which provides more than 800 input/output (I/O). The flip-chip technique started from solder bumps with a diameter range of 75-200 $\mu$m, while copper-pillar interconnection with solder cap nowadays reaches down to the diameter less than 40 $\mu$m. The copper-pillar technique not only provides high-density I/O, but also does it offer lower conductive loss. 

The evolution of IC-embedding or fan-out packaging has been recently imperative. Embedded wafer-level ball grid array (eWLB) pioneered by Infineon and fan-out wafer-level packaging (FOWLP) are drawing attention as mm-wave packages. This embedding technology eliminates the use of wire bonding which not only introduces significant high-frequency loss and parasitics, but also increases the footprint for high-pin count die. Transceiver ICs are embedded in a reconfigured molded wafer with compression molding process. Multiple redistribution layers (RDLs) are formed to fan-out the baseband signals, and through-mold vias are utilized for vertical interconnections. From the industry standpoint, Infineon, TSMC, ASE, Amkor, Deca Technologies, JCET, and several more companies have been leading the wafer-level fan-out technology for RF/mm-wave applications. In conjunction with the molding-compound-based waver-level IC-embedding technology, laminate-based and glass-based panel-level embedding technology is also emerging as an alternative candidate, mainly led by Samsung, Unimicron, ASE, and Deca Technologies from the industry, and Institute of Microelectronics (IME) \cite{2020Che} and Georgia Tech Packaging Research Center (GT-PRC) from the research standpoint \cite{2018Watanabe,2019Ravichandran}.

\begin{figure}[t]
		\centering
		\includegraphics[width=0.4\textwidth]{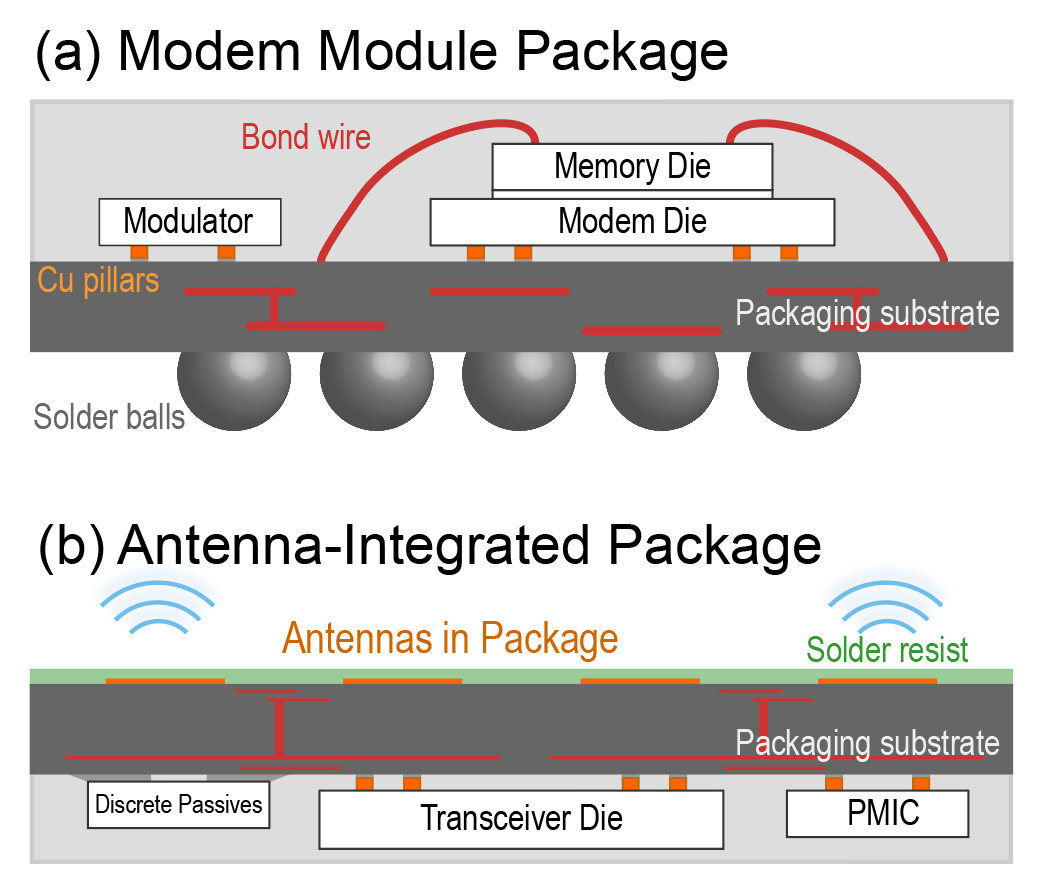}
    	\caption{Examples of system-integrated mm-wave packages (a) modem SoP (b) antenna-integrated package.}
		\centering
		\label{fig:modem_aip}
	\end{figure}

Fig. \ref{fig:modem_aip} illustrates the key modem and antenna-integrated packages. Most cases are such modules employ the combination of multiple techniques from the three interconnection methods. Bond wires could serve as interconnection between the PCB board and memory that is stacked atop the logic or modem die, while the modem die with high pin count entails flip-chip interconnects to provide clean signal and mitigated signal delay. The most popular assembly method to integrate a mm-wave phased antenna array with ICs interconnected is, as of now, the flip-chip technique because of the process cost and supply-chain maturity. Conductive materials are selected from copper pillar or C4 bumps, depending on the pin count and sensitivity of assembled dies to the conductive loss caused by interconnections. 

\subsection{Materials for Core, Prepreg, and Buildup}
\label{subsec:material}
The selection of dielectric materials is critical to obtain the desired electrical performance of ICs and antenna arrays integrated in 5G module packages. A wide variety of dielectric materials are studied and investigated to meet requirements for various applications such as handset devices, radar modules, base-stations, and satellites. Well-designed dielectric materials provide improved link budget, low signal dissipation, high signal or power density, desired frequency response of passive components, small footprint of elements, lower module thickness, high antenna efficiency and EIRP, beam width, angular coverage, energy consumption, and miniaturization of antennas.

LTCC has been employed for high-frequency antenna modules such as WiGig, alternatively known as 60 GHz WiFi, which includes IEEE 802.11ad standard and also the upcoming IEEE 802.11ay standard. The large number of metal layers and relatively-low tan $\delta$ are the major advantages of LTCC, as indicated in Table \ref{tab:core}. Its low coefficient of thermal expansion (CTE) and high thermal conductivity also lead to high reliability and robustness to temperature variations. The major challenges are the low density of signal routings in internal layers and size scalability. As LTCC substrates are fabricated with screen-printing and co-firing processes, the features of RDL are generally large ($> 125\: \mu$m) due to the dimensional limit of screen-printing masks and alignment accuracy between layers. This large feature size results in low density of signal routing and entails high number of metal layers, resulting in modules.

    \begin{table}[b]
	\centering
	\caption{Low-loss dielectric materials recently used for 5G substrate technology}
    \begin{tabular}{c|c|cc|c}
\hline
                                              & \multirow{2}{*}{Materials}            & Dk                  & Reported    & Major            \\
                                              &                                       & Df ($\times ^{-4}$) & frequency   & suppliers        \\ \hline \hline
\multirow{12}{*}{\rotatebox{90}{Organic}}     & Bismaleimide                          & 3.4                 & 10 GHz      & Mitsubishi       \\
                                              & Triazine                              & 40 – 50             &             &                  \\
                                              & (BT)                                  &                     &             &                  \\ \cline{2-5} 
                                              & Polyphenyl-                           & 3.25 – 3.4          & 1 – 50 GHz  & Panasonic,       \\
                                              & ethers                                & 20 – 50             &             & Risho Kogyo      \\
                                              & (PPE)                                 &                     &             &                  \\ \cline{2-5}
                                              & Liquid-crystal                        & 2.9                 & 10 GHz      & Rogers,          \\
                                              & Polymer                               & 25                  &             & Murata           \\
                                              & (LCP)                                 &                     &             &                  \\ \cline{2-5}
                                              & Polytetrafluoro-                      & 2.2                 & 10 GHz      & Rogers,          \\
                                              & ethylene                              & 9                   &             & DuPont           \\
                                              & (PTFE)                                &                     &             &                  \\ \hline
\multirow{9}{*}{\rotatebox{90}{Inorganic}}    & Low-temp.                             & 6                   & 60 GHz      & Hitachi Metals,  \\
                                              & cofired ceramic                       & 18                  &             & Kyocera, TDK     \\
                                              & (LTCC)                                &                     &             &                  \\ \cline{2-5}
                                              & Borosilicate                          & 5.4                 & 10 – 60 GHz & AGC,             \\
                                              & glass                                 & 50 – 90             &             & Corning, Schott, \\
                                              &                                       &                     &             & 3DGS, NSG        \\ \cline{2-5}
                                              & Fused silica                          & 3.8                 & 10 – 60 GHz & AGC,             \\ 
                                              &                                       & 3 – 4               &             & Corning, Schott, \\
                                              &                                       &                     &             & 3DGS, NSG        \\ \hline
    \end{tabular}

    \label{tab:core}
    \end{table}

The mainstream platform for antenna integrated packages is still based on multi-layered organic substrates. The process for the multi-layered organic substrates for mm-wave modules is similar to that of PCB manufacturing and is cost-effective due to the compatibility with the existing supply chain and high-volume manufacturing for consumer electronics. The organic materials are designed to show lower tan $\delta$, compared to the traditional FR4, where tan $\delta$ is higher than 0.02 in the mm-wave frequency bands. The mainstream copper-clad laminates (CCL) and prepregs used for formation of multi-layered organic substrates comprises of four classes of polymers: 1) Bismaleimide Triazine (BT), 2) Polyphenyleneether (PPE), 3) Liquid-crystal Polymer (LCP), and 4) Polytetrafluoroethylene (PTFE), as listed in Table \ref{tab:core}. Unlike glass-cloth epoxy resin (e.g., FR4), glass-cloth PPE substrates feature high glass-transition temperature (Tg), low water absorption, low dielectric constant (both Dk and Df). PPE-based substrates are typified by MEGTRON6 (core and prepreg) from Panasonic, and CS-3376C from Risho Kogyo. Similar to PPE resin, the lamination of BT-based core with build-up dielectrics form the multi-layered organic substrates. BT-based laminates provide low CTE, low shrinkage, and high peel strength with copper. LCP and PTFE are gaining attention because of low tan $\delta$ at the mm-wave frequency range. Unlike PPE and BT, these materials require high-temperature and high-pressure processes for lamination and thermal compression. Halogenated substrates are not usually preferred for handset applications for environmental and safety concerns. 

The mechanical and process limitations of organic laminates from their low modulus and high CTE are addressed with emerging inorganic substrates, as listed in Table \ref{tab:core}. Glass substrates offer a wide range of Dk (3.7 -- 8) and Df (0.0003 for fused silica to 0.006 for alkaline-free borosilicate), smooth surface, good dimensional stability ($< 2 \mu$m for 20-mm substrates), large-panel scalability, ability to form  fine-pitch through vias, stability to temperature and humidity, and tailorable CTE depending on the packaged components \cite{2019Hayashi,2019Martin,2019Bowrothu,2018Watanabe,2020Xia}. Companies such as Samtec and Unimicron manufacture glass-based packages. However, glass-based packaging still poses challenges associated with process immaturity and higher cost resulting from lack of the supply-chain readiness and the nature of glass such as brittleness or robustness and difficulty in handling.  

In conjunction with the substrate materials, the formation of RDL and microvias \cite{2020Watanabe_cpmt_via} becomes more critical to form high-density interconnects ($< 10\: \mu$m) and to enable tight integration of components. Dry-film dielectric materials show good compatibility with panel-level packages with double-side RDLs and offer a wide thickness variation ($> 5\: \mu$m). Low-loss or low tan $\delta$ materials are developed in the industry for recent mm-wave packages, as shown in Table \ref{tab:dryfilm}. Low-loss dry-films, which are typically epoxy-based, usually contain silica or ceramic fillers to lower the dissipation factor (tan $\delta$). The fillers increase Young’s modulus, decrease CTE, and lead to technical challenges such as the narrow process window, poor adhesion to substrate materials, and reliability in harsh environment. The other method to form dielectric is the use of liquid-based dielectric such as polyimides. As opposed to dry films, liquid-based dielectric materials are more compatible with wafer processes or build-up layers on one side of the substrate. Non-filler photo-sensitive or photo-imageable low-loss liquid-based dielectric materials designed for 5G mm-wave applications are emerging to form fine features below 5 $\mu$m \cite{2019Ito,2014Connor,2018Tomikawa}. Low-loss liquid-based dielectric materials from major suppliers are listed in Table \ref{tab:liquid}. These materials are, however, not still in high volume manufacturing phase. The other challenge is that these materials are spin-coated or slit-coated and have difficulty in forming films with thickness more than 40 $\mu$m.

    \begin{table}[t]
	\centering
	\caption{Low-loss build-up dry films for high-density signal routings for high-frequency applications}
    \begin{tabular}{cccccc}
    \hline
                         & Dk   & Df                    & Frequency & CTE     & $T_g$  \\
                         &      & ($\times 10^{-4}$)    &           & (ppm/K) & ($^{\circ}$C) \\ \hline \hline
        Ajinomoto        & 3.3  & 44      & 5.8 GHz     & 20        & 153 \\
        DOW              & 2.57 & 32      & 1 MHz       & 63        & 250 \\
        Hitachi \par Chem.    & 3.3  & 34      & 5 GHz       & 17        & 233 \\
        Sekisui          & 3.3  & 37      & 5.8 GHz     & 27        & 183 \\
        Taiyo Ink        & 3.3  & 25 – 30 & 5 – 60 GHz  & 20        & 160 \\ \hline
    \end{tabular}
    \label{tab:dryfilm}
    \end{table}
    
    \begin{table}[t]
	\centering
	\caption{Low-loss photosensitive (photo-imagable) dielectrics for high-density signal routings for high-frequency applications}
    \begin{tabular}{cccccc}
    \hline
                         & Dk   & Df                 & Frequency    & Min.      & Elonga-       \\
                         &      & ($\times 10^{-4}$) &              & L/S       & tion (\%)     \\ \hline \hline
        DOW              & 2.65 & 8                  & $<$20 GHz    & 18 $\mu$m & 8             \\
        Hitachi Chem.    & 2.4  & 18                 & 10 GHz       & --        & --            \\
        JSR              & 2.6  & 48                 & $<$ 40 GHz   & 8 $\mu$m  & $>$ 50        \\
        Toray            & 2.9  & 30                 & 1 GHz        & 30        & –             \\ \hline
    \end{tabular}
    \label{tab:liquid}
    \end{table}

\subsection{Integration of Passive Components}
\label{subsec:passive}
Passive components play a key role in wireless system implementations as there is a need to provide matching impedances for components such as PAs and LNAs, filtering, tuning and biasing \cite{RN121}. Passive components also make up functions such as couplers, baluns, power combiners and dividers, filters, phase shifters, circulators and isolators. Duplexers are usually paired with RF ICs for their nominal operation on the system level. It is typically estimated that passive components account for 90\% of the component count, 80\% of the size and 70\% of the cost \cite{RN107}.

\subsubsection*{Discrete Lumped Circuits for sub-6 GHz 5G bands}
\begin{figure}[b]
	\centering
	\includegraphics[width=0.48\textwidth]{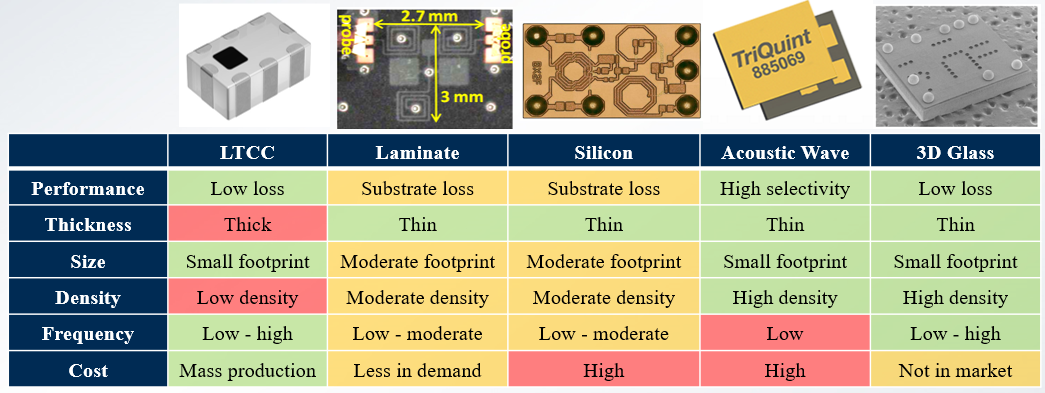}
    \caption{Evolution of passive component technologies (Courtesy of Zihan Wu, Finisar Corporation)}
	\centering
	\label{fig:Passive_comp_review_Zihan}
\end{figure}

\begin{figure}[t]
	\centering
	\includegraphics[width=0.4\textwidth]{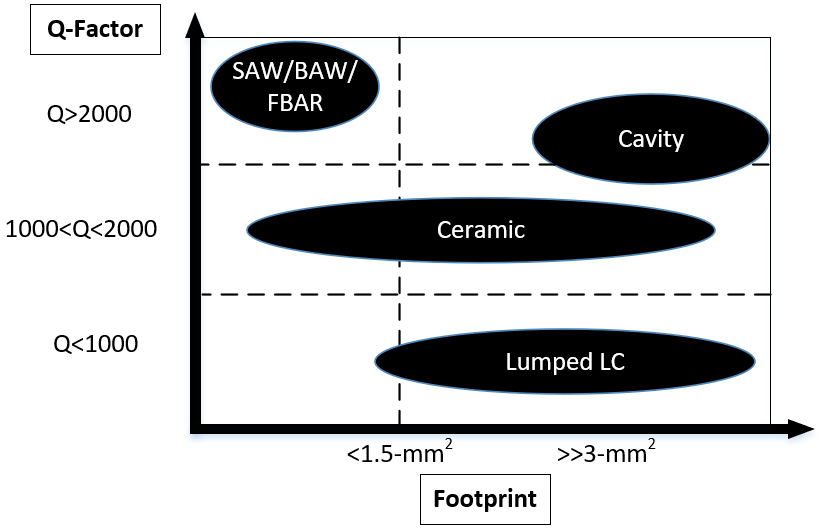}
    \caption{Comparison of various filter technologies at 1 GHz in terms of performance (Q-factor) and footprint)}
	\centering
	\label{fig:compare_filter_techn_at_1GHz_Qvsfootprint}
\end{figure}

The evolution of passive components started with low temperature co-fired ceramic (LTCC), which became very popular as the surface-mount technology (SMT) components for resistors, inductors and capacitors. 
A comparison of five different substrate technologies for passive components in terms of performance, thickness, size and some other parameters is given in Fig. \ref{fig:Passive_comp_review_Zihan} \cite{Zihan_thesis}. 
Moreover, a comparison of filter performance (Q-factor) vs. footprint is given in Fig. \ref{fig:compare_filter_techn_at_1GHz_Qvsfootprint}. Acoustic wave technologies have surpassed the typical lumped and distributed LC networks in sub-6 GHz range in realizing high performance filters, resonators, oscillators and delay lines, and have found many applications in 4G and LTE networks \cite{Qorvo_RF_Filter_for_dummies}.

\subsubsection*{Distributed Components for mm-wave}
Over the past few decades, several theoretical advancements have been made in the design and fabrication of planar passive components such as filters, power dividers, couplers and baluns with their evolution in terms of reduction in size while improving performance metrics \cite{Approach_5G_mmwave_filter_challenge}. Some of the transmission lines of choice are microstrip, conductor-backed coplanar waveguide (CBCPW) and substrate-integrated waveguide (SIW). The technologies include LTCC, organic laminates, state-of-the-art integrated fan-out wafer-level packaging (InFO WLP) and ultra-thin laminated glass for passive components realization. A brief review of prior art of these components is this section.  

LTCC facilitated initial advances in mm-wave distributed components because of its ability to integrate complex 3D multilayered conductor patterns with through and blind vias. A 4-pole dual-mode resonator filter on LTCC for 30 GHz center frequency with FBW of 4.67\% and an insertion loss of 2.95 dB using two transmission zeros is reported on a five-metal layer stackup in \cite{RN151}. Similarly, two-pole, two-stage SIW single cavity filter with embedded planar resonators realized on five layers of LTCC with center frequency of 28.12 GHz, 15\% FBW and 0.53 dB insertion loss is reported for 5G applications in \cite{LTCC_example_fig}. A four-pole, four-cavity SIW filter with 2.66 dB insertion loss at the center frequency of 27.45 GHz with 3.6\% FBW is designed on nine layers of LTCC in \cite{LTCC_SIW_thesis}. 60 GHz band filters have also been of interest in academia using LTCC stackups \cite{Guo_microwave_and_millimeter}.

{
\begin{figure}[b]
	\begin{minipage}{.5\linewidth}
	\centering
	\subfloat[]{\label{fig:organic_filter_example_a}\includegraphics[width=1.75in]{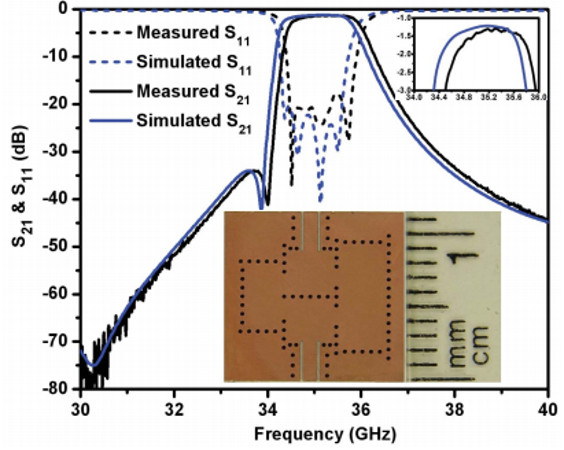}}
	\end{minipage}%
	\begin{minipage}{.5\linewidth}
	\centering
	\subfloat[]{\label{fig:organic_filter_example_b}\includegraphics[width=1.75in]{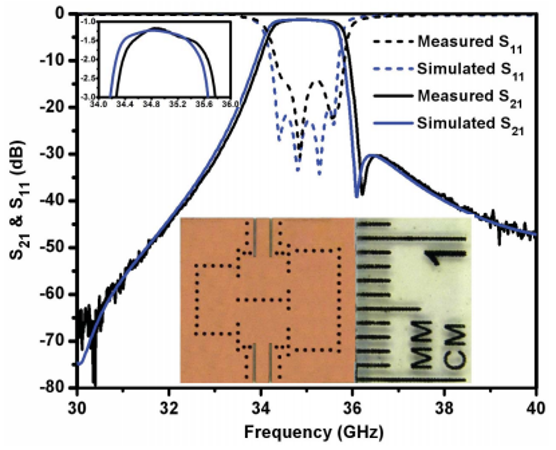}}
	\end{minipage}\par\medskip
	\caption{Ka-band 4-pole SIW filter on organic laminate (a) TZ on left of passband, and (b) TZ on right of passband \cite{organic_example_filter}}
	\label{fig:organic_filter_example}
\end{figure}
}

Although LTCC became a very attractive option for fabricating microwave and mm-wave passive components for a couple of decades, challenges such as complexity and cost of fabrication due to high number of layers, loss at mm-wave frequencies and surface roughness led to developments in the wake of relatively new planar organic substrates with stable electrical properties and ease of fabrication. Chen \textit{et al.}, have reported Ka-band four-pole bandpass filters on SIW having asymmetric frequency response for diplexer applications for high rejection between neighboring channels \cite{organic_example_filter}. The filters are fabricated on Rogers RT/Duroid 6002 substrate with thickness of 0.508 mm by using a low-cost PCB process and have transmission zeros (TZ) on either side of the passband as shown in Fig. \ref{fig:organic_filter_example}. The center frequency of these filters is 35 GHz with 3.7\% FBW and insertion loss of 1.25 dB. Similar filter structures using ridge gap waveguide (RGW) for 30.5 GHz center frequency are demonstrated by Sorkherizi \textit{et al.} in \cite{Sorkherizi_self-packaged_lowloss}. Another research focused on microstrip coupled-line bandpass filters with an insertion loss of 4 dB and 4.67\% FBW centered at 27.85 GHz on Teflon whereas the filter fabricated on Alumina shows 3 dB insertion loss and 7.8\% FBW centered at 38.5 GHz \cite{4140571}. Tsai \textit{et al.} at TSMC have realized high-performance passive devices for mm-wave system including inductor, ring resonator, power combiner, coupler, balun, transmission line and antennas using InFO WLP \cite{InFO_WLP_Passives}. The inductors have a Q factor of over 40 and the power combiner, coupler and balun show lower transmission loss than on-chip passives. The schematic and fabricated passive devices are shown in Fig. \ref{fig:InFO_WLP_Passives}. 

{
\begin{figure}[t]
	\begin{minipage}{.5\linewidth}
	\centering
	\subfloat[]{\label{fig:InFO_WLP_Passives_a}\includegraphics[width=1.75in]{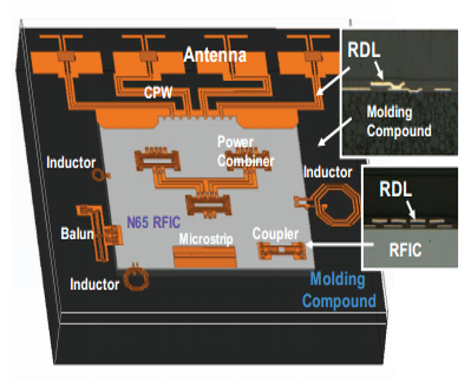}}
	\end{minipage}%
	\begin{minipage}{.5\linewidth}
	\centering
	\subfloat[]{\label{fig:InFO_WLP_Passives_b}\includegraphics[width=1.5in]{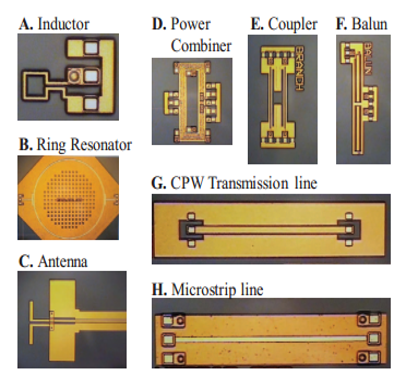}}
	\end{minipage}\par\medskip
	\caption{(a) Schematic of mm-wave circuit including RFIC, passive devices and antennas (b) realized passive devices \cite{InFO_WLP_Passives}}
	\label{fig:InFO_WLP_Passives}
\end{figure}
}

{
\begin{figure}[b]
	\begin{minipage}{.5\linewidth}
	\centering
	\subfloat[]{\label{fig:InFO_WLP_PD_a}\includegraphics[width=1.75in]{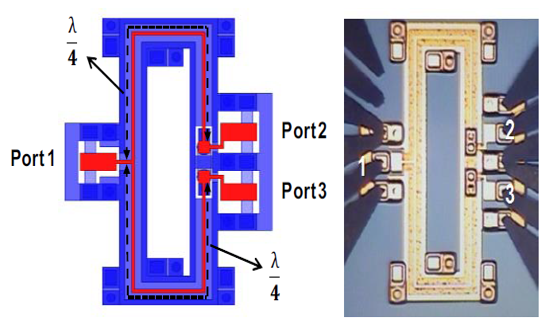}}
	\end{minipage}%
	\begin{minipage}{.5\linewidth}
	\centering
	\subfloat[]{\label{fig:InFO_WLP_PD_b}\includegraphics[width=1.5in]{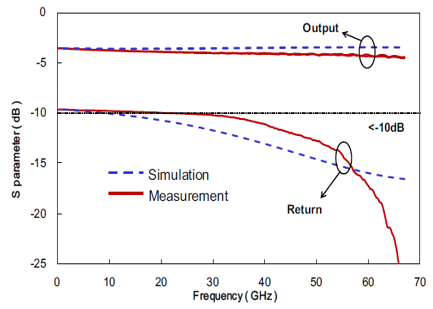}}
	\end{minipage}\par\medskip
	\caption{CPW power divider on InFO RDL (a) design schematic and fabricated layout, and (b) s-parameters \cite{InFO_WLP_PD}}
	\label{fig:InFO_WLP_PD}
\end{figure}
}

Beyond filters, extensive work is also reported on integrated power dividers for 5G antenna-array applications. A key innovation in these power dividers lies in performance breakthroughs, physical configurations and functional integration with wafer fan-out packaging. Transmission line losses of 0.35 and 0.34 dB/mm are achieved with CPW and microstrip loss respectively. One such advance presented by Hsu \textit{et al.} is an equal-split ultra-miniaturized (960$\times$360 \(\mu\)m\textsuperscript{2}) CPW power divider on integrated fanout (InFO) redistribution layers (RDL), with excellent ground shielding capability along with other passive components such as balun and coupler \cite{InFO_WLP_PD}. The power divider exhibits a 10-dB return loss from DC-67 GHz with insertion loss of 4.3 dB from 30-67 GHz. The layout of the power divider with its s-parameters is shown in Fig. \ref{fig:InFO_WLP_PD}. 

\begin{figure}[t]
	\centering
	\includegraphics[width=0.48\textwidth]{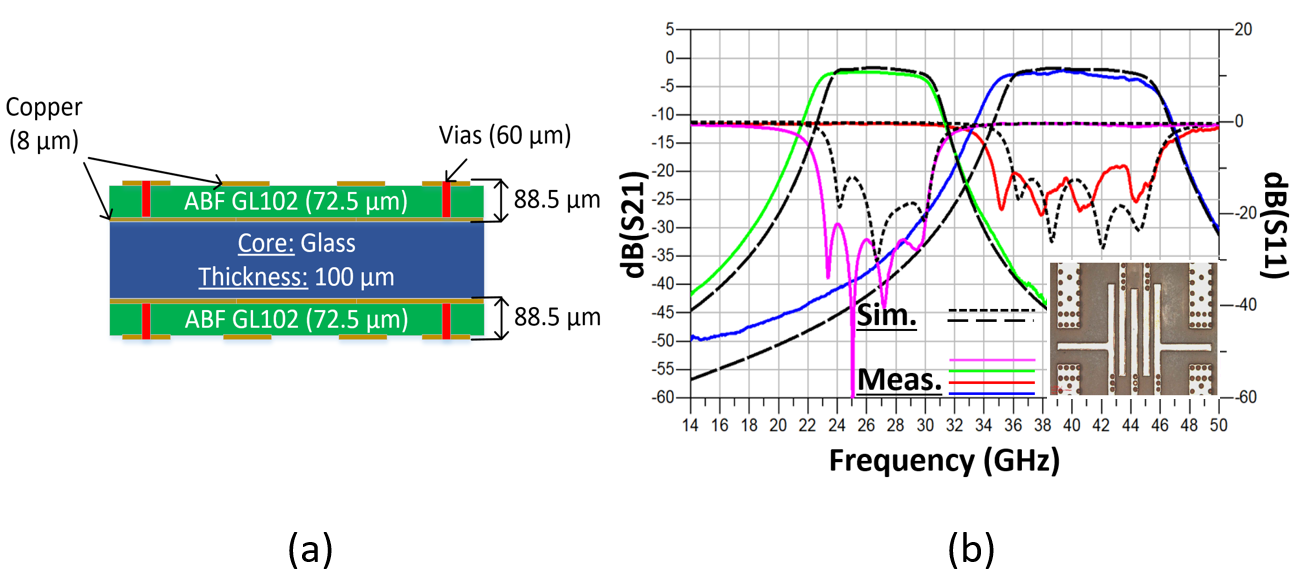}
	\caption{(a) Ultra-thin laminated glass-based stackup for 5G filters, (b) s-parameters of 5G 28 and 39 GHz interdigital filters and a fabricated interdigital filter coupon \cite{Filters_ECTC2018_Ali, Filters_MWCL2018_Ali}}
	\centering
	\label{fig:glass-based-filters}
\end{figure}

Various miniaturized, high performance passive components have been demonstrated on glass \cite{ECTC2017}. GT-PRC has developed key building blocks such as interconnects, antennas and filters using laminated glass substrate. Test vehicles with six metal layers with the semi-additive patterning (SAP) process enables higher precision on RDL, less than 2\% variance in dimensions and insertion loss of 0.1 dB/mm with 15 \(\mu\)m buildup dielectric and 0.05 dB/mm with 75 \(\mu\)m dielectric \cite{2018Watanabe}. Package-integrated and ultra-thin lowpass and bandpass filters with footprint smaller than 0.5$\lambda$\textsubscript{0}$\times$0.5$\lambda$\textsubscript{0} at the operating frequencies of 28 and 39 GHz bands are developed for 5G and mm-wave small-cell applications as shown in Fig. \ref{fig:glass-based-filters} \cite{Filters_ECTC2018_Ali, Filters_MWCL2018_Ali}. Such package integration of 5G filters with ultra-short 3D interconnects allow for low interconnect losses that are similar to that of on-chip filters, but low component insertion loss of off-chip discrete filters. These thin-film filters exhibit a cross-sectional height of less than 200 \(\mu\)m and can be utilized either as embedded components or IPDs in module packages and they can be configured as diplexers as well \cite{Diplexers_ECTC2020_Ali}. A variety of filter structures based on different substrate technologies are compared in Table \ref{tab:comparison_filters_in_research}.

{
\begin{table}[b]
	\centering
	\caption{Comparison of filters using various substrate technologies for 5G mm-wave applications}
	\label{tab:comparison_filters_in_research}
	\renewcommand{\arraystretch}{1.5}
	\resizebox{\linewidth}{!}{%
		\begin{tabular}{ccccccc} 
			\hline
			\textbf{Ref.} & \begin{tabular}[c]{@{}c@{}}\textbf{Substrate Technology /} \\\textbf{Structure} \end{tabular}  & \textbf{Order} & \begin{tabular}[c]{@{}c@{}}\textbf{IL} \\\textbf{(dB)} \end{tabular} & \begin{tabular}[c]{@{}c@{}}\textbf{f\textsubscript{c}}\\\textbf{(GHz)} \end{tabular} & \begin{tabular}[c]{@{}c@{}}\textbf{BW}\\\textbf{(GHz)} \end{tabular} & \begin{tabular}[c]{@{}c@{}}\textbf{Size} \\\textbf{(mm\textsuperscript{2})} \end{tabular}  \\ 
			\hline\hline
			\cite{LTCC_example_fig}     & LTCC / SIW                   & 2     & 0.53                                               & 28.12                                               & 4.29                                               & 33.5                                                   \\ 
			\hline
			\cite{LTCC_SIW_thesis}     & LTCC / SIW                   & 4     & 2.66                                               & 27.45                                              & 0.98                                               & 43.5                                                   \\ 
			\hline
			\cite{RN151}     & LTCC / Cavity                & 4     & 2.95                                               & 30                                                 & 1.4                                                & 88                                                     \\ 
			\hline
			\cite{organic_example_filter}     & Rogers Laminate / SIW            & 4     & 1.25                                               & 35                                                 & 1.3                                                & 121                                                    \\ 
			\hline
			\cite{4140571}     & Alumina / Microstrip          & 3     & 3                                                  & 38.5                                              & 3                                                & 7.1                                                      \\ 
			\hline
			\cite{8059037}    & \begin{tabular}[c]{@{}c@{}}Rogers Laminate/\\Air-filled SIW \end{tabular} & 4     & 3.9                                                & 21                                                 & 0.23                                               & 746                                                    \\ 
			\hline
			\cite{Filters_MWCL2018_Ali}     & Glass / Microstrip           & 5     & 2.6                                                & 27                                                 & 5                                                  & 6.9                                                    \\ 
			\hline
			\cite{Filters_MWCL2018_Ali}     & Glass / Microstrip           & 5     & 1.43                                               & 40.25                                              & 6.5                                                & 5.5                                                    \\
			\hline
		\end{tabular}
	}
\end{table}
}

In the discussion above, the passive components and their integration is only discussed from the perspective of published research. Component manufacturers have advanced discrete passive components such as filters, couplers and dividers. A comparison of various 5G filters from industry is given in Table \ref{tab:comparison_filters_commercial}.

{
\begin{table}[t]
\caption{Comparison of commercial filters for 5G mm-wave applications}
\label{tab:comparison_filters_commercial}
\centering
\renewcommand{\arraystretch}{1.5}
\resizebox{\linewidth}{!}{%
\begin{tabular}{ccccc} 
\hline
\textbf{Company \& Ref.}     & \begin{tabular}[c]{@{}c@{}}\textbf{IL}\\\textbf{(dB)}\end{tabular} & \begin{tabular}[c]{@{}c@{}}\textbf{f\textsubscript{c}}\\\textbf{(GHz)}\end{tabular} & \begin{tabular}[c]{@{}c@{}}\textbf{BW}\\\textbf{(GHz)}\end{tabular} & \begin{tabular}[c]{@{}c@{}}\textbf{Size}\\\textbf{(mm\textsuperscript{2})}\end{tabular}  \\ 
\hline\hline
TDK  \cite{tdk2019}             & 1                                                                  & 28.5                                                                & 2                                                                   & 5                                                                      \\ 
\hline
Knowles (band n257) \cite{knowles2019_bandn257} & 3.5                                                                & 28                                                                  & 3                                                                   & 12.65                                                                  \\ 
\hline
Knowles (band n258) \cite{knowles2019_bandn258} & 2.9                                                                & 25.875                                                              & 3.25                                                                & 12.65                                                                  \\ 
\hline
Knowles (band n260) \cite{knowles2019_bandn260} & 2.5                                                                & 38.5                                                                & 3                                                                   & 14.19                                                                  \\
\hline
Pasternack\textsuperscript{*} \cite{Pasternack_filter} & 2.5                                                                & 29.25                                                                & 3.5                                                                   & 416.7                                                                  \\
\hline
\end{tabular}
}
\begin{tablenotes}
	\item[1] \footnotesize *9-section bandpass filter. 
\end{tablenotes}
\end{table}
}

\subsection{Antenna Systems in Package (AiP)}
\label{subsec:aip}
The 5G wireless communication system entails highly-integrated radio access solutions, incorporating advanced phased-array antenna and transceiver front-end technology to support high radiated power, large signal-to-noise ratio, as well as beamforming, scanning in elevation and azimuth directions within a wide range \cite{2015Gupta,2019Gu}. As the antenna element size and pitch are correlated with the wavelength, antenna-in-package solutions become more feasible at mm-wave frequencies unlike discrete antennas for 4G/LTE. Major package-integrated antenna structures are patch antennas and dipole antennas. More number of antenna elements ($N$) results in additional gain ($10 \log N$). A  2 $\times$ 2 antenna array in Qualcomm QTM052 is reported to have a gain of 20 dBi, which comprises of 5 dBi per element, 6 dBi from summation, 6 dBi from array formation and additional 3 dBi gain from single polarization.

Patch antennas feature the main lobe in the elevation direction and enable dual-polarization which increases the channel capacity. The size of a patch-antenna array is approximately $\lambda/2$ (effective half wavelength on dielectric) with a pitch of $\lambda_0/2$ (half-wavelength in air). These fundamental design rules lead to the requirement of high-Dk materials for miniaturization. High-Dk materials, however, degrade the bandwidth and gain. The thickness or separation of the patch and ground is critical since the antenna bandwidth increases almost linearly with the thickness. Another technique to increase bandwidth is the stacked-patch antenna. The stacked patches create two resonances at different frequencies, which also calls for high accuracy of layer-to-layer alignment to obtain desired frequency responses. Antenna designs are performed considering the trade-off between the size and performance (i.e., bandwidth and gain). Common feeding methods are the via-feed or aperture-coupling feed. Via-feeding is relatively simple and guarantees the accuracy of feeding points into the antenna patch. However, the challenge is the limited bandwidth because of the inductive reactance caused by the via \cite{2019Liu}. Conformal via shielding is employed to emulate coaxial feed to reduce the undesired inductance and enable better impedance matching. In contrast, aperture-coupled feed leads to higher bandwidth than the via feed, while the layer-to-layer alignment is extremely critical; a few-micron shift between layers causes the frequency shift and may result in not covering the targeted frequency bands. 

Dipole antennas are employed for covering the azimuth direction with a single polarization. Since the length of the dipole is nearly $\lambda_0/2$, the design process is relatively simple. Dipole antennas generally offer wide bandwidths, compared to patch antennas. The gain is controlled by placing additional number of directors in a modified antenna topology referred to as Yagi-Uda antenna. Dipole antenna and Yagi-Uda antenna are fed by two differential transmission lines to reverse the phases of the provided signals. Co-planar-waveguide and stripline signal routing are more common in multi-layered packages because of their inherent shielding features and minimal cross-talk with other nearby transmission lines, components, and antennas.

\begin{figure}[b]
		\centering
		\includegraphics[width=0.45\textwidth]{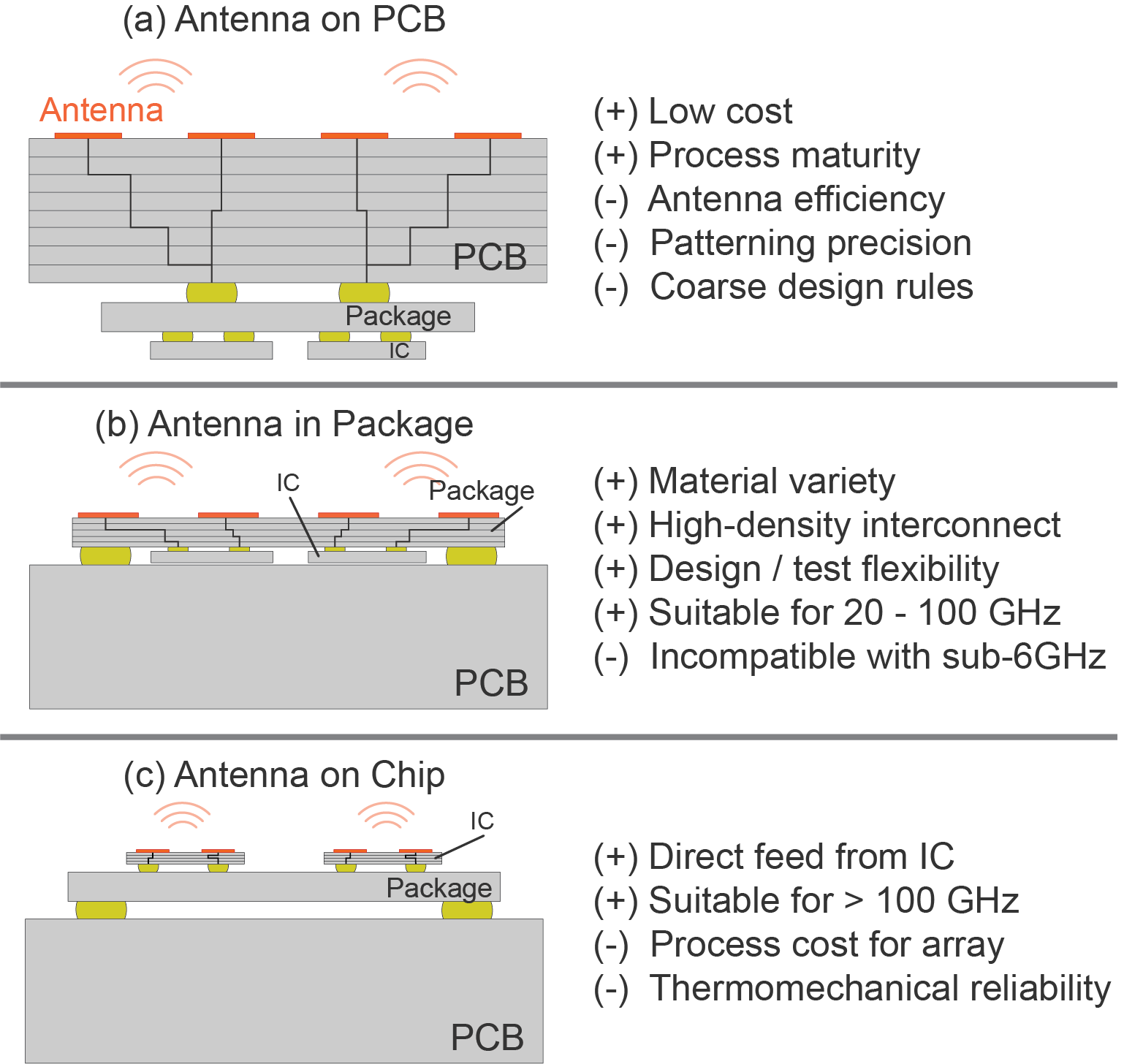}
    	\caption{Three approaches for mm-wave antenna implementation. (a) Antennas on PCB, (b) antennas in package, and (c) antennas on chip or wafer.}
		\centering
		\label{fig:aips}
	\end{figure}

There are several options to implement mm-wave antenna array, as shown in Fig. \ref{fig:aips}. The first option refers to implementing antennas directly on the PCB board (Fig. \ref{fig:aips}a) \cite{2015Shahramian,2018Kibaroglu,2019Thai}. The most notable advantage is the lower cost than the other options because of the supply-chain maturity. The major challenges of antenna on PCB are manufacturing process and tolerances. The relatively-coarse design rules do not allow designers to layout fine structures in transmission-line widths, spaces, via diameters and pitch, and accurate layer-to-layer alignment. Fig. \ref{fig:aips}b illustrates the most viable option for 5G mm-wave applications, the implementation of AiP. In contrast, Fig. \ref{fig:aips}c shows the direct antenna implementation onto the IC wafer. Although the antenna-on-chip approach offers the lowest feed-line loss and parasitics and provides direct integration with other front-end circuitry \cite{2018Li}, the challenges include antenna efficiency, process scalability for large array, yield and cost, thermomechanical reliability issues, and design flexibility.

\section{Current Development in Heterogeneous Package Integration for 5G}
\label{sec:demo}

\begin{figure}[t]
		\centering
		\includegraphics[width=0.48\textwidth]{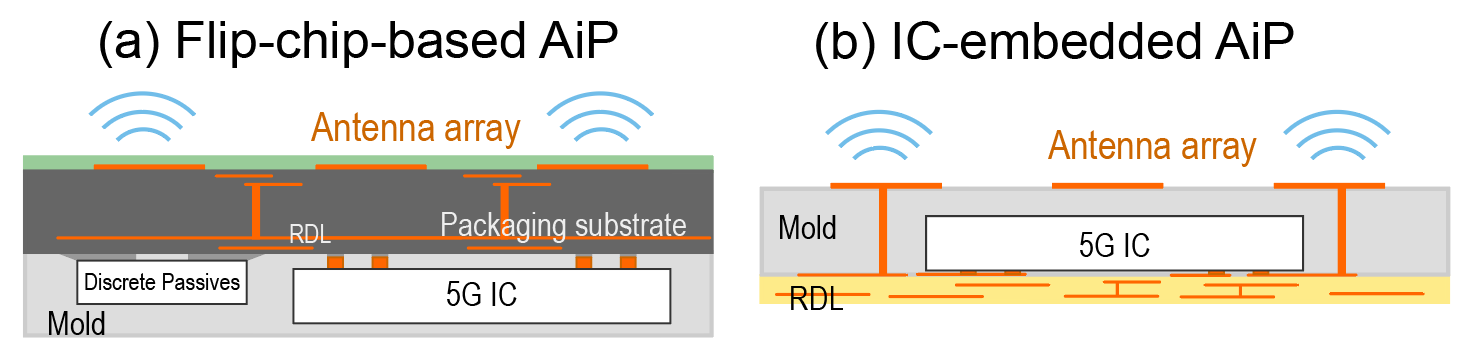}
    	\caption{Packaging structures for antenna integration (a) currently-viable flip-chip-based configuration (b) IC-embedded configuration.}
		\centering
		\label{fig:fc_fo}
	\end{figure}

This section introduces heterogeneous package integration for mm-wave antenna-integrated packages. Recent mm-wave antenna-integrated packages are classified into two categories based on interconnection techniques: a) chip-last or flip-chip structures, as shown in Fig. \ref{fig:fc_fo}a and b) chip-embedded structures exemplified in Fig. \ref{fig:fc_fo}b. System-level architectures and applications are discussed along with the packaging structures and materials  in Section \ref{subsec:material}.

	    \begin{figure}[b]
		\centering
		\includegraphics[width=0.48\textwidth]{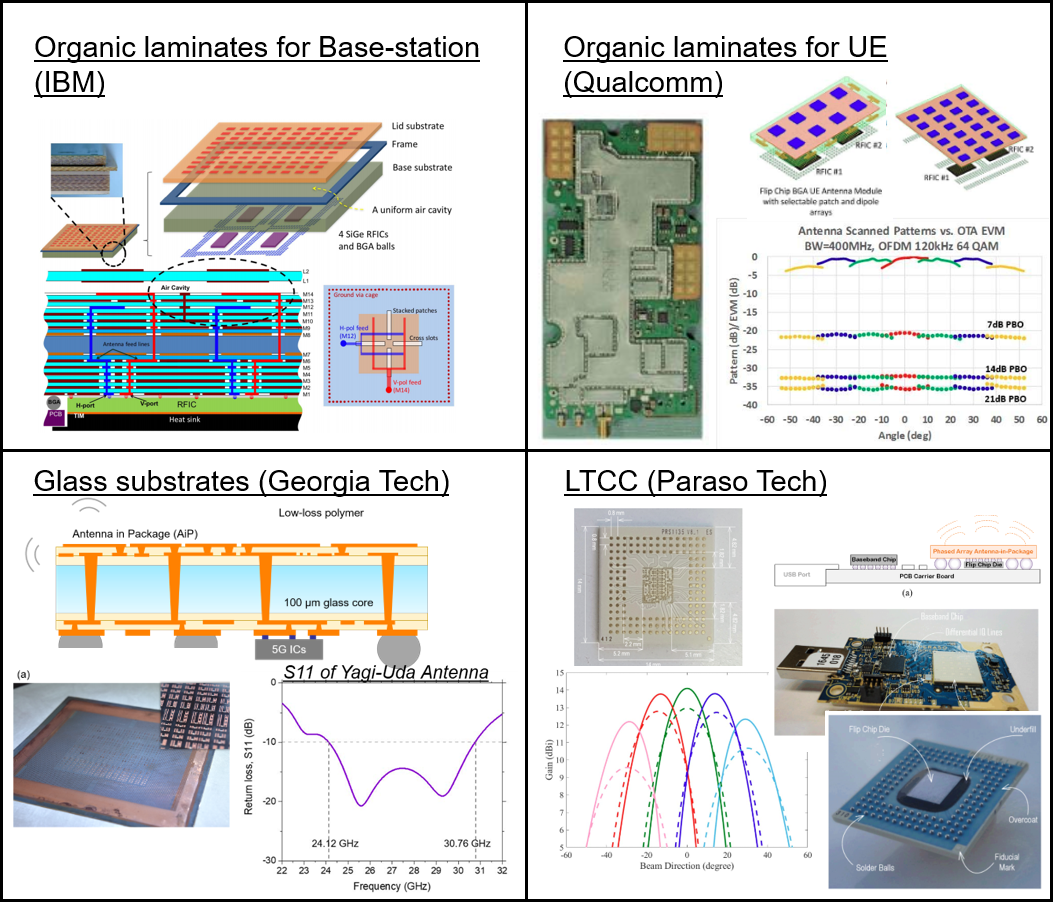}
    	\caption{Demonstrated mm-wave modules with flip-chip-based AiP: organic laminates from IBM \cite{2019Gu} and Qualcomm \cite{2018Dunworth}, glass-based substrates \cite{2018Watanabe}, and LTCC \cite{2019Rashidan}.}
		\centering
		\label{fig:flipchip}
    \end{figure}

IBM made pioneering advances in AiP with organic laminate substrates for base stations, as shown in Fig. \ref{fig:flipchip}. The multi-chip antenna-in-package includes 64-array embedded antennas with dual-polarized operation in Tx and Rx modes, with four transceiver ICs that are assembled on the backside with flip-chip technology. In addition, to ensure thermal management, a heat sink is added to ball-grid array (BGA) interface, which allowed to realize consistent board-level assembly. This module operates at 28 GHz and achieves more than 50 dBm EIRP in Tx mode and $\pm 40^{\circ}$ scanning range with 70$\times$70$\times$2.7 mm$^3$ \cite{2019Gu,2017Gu}. Ericsson developed similar AiP to implement a reference design for base stations. UCSD also reported 64 dual-polarized dual-beam single-aperture 28-GHz phased array for 5G MIMO, which is implemented in low-cost organic laminates. A 2$\times$4 dual-beam former chip is flip-chip assembled on the other side of the antenna array. The array achieved an EIRP of 52 dBm at 29 GHz and scanned $\pm$ 50$^{\circ}$ and $\pm$ 25$^{\circ}$ in azimuth and elevation planes, respectively, with cross-polarization rejection $>$ 30 dB \cite{2019Nafe}.

Intel \cite{2019Thai} has reported a stack-up that consists of an organic laminate-prepreg core for antenna implementation, trace routing, a flip-chip RFIC module that is responsible to quad-feed the antenna where each port is responsible for one band and polarization, and BGA for connection to the main PCB, along with conformal shielding. The package size is 30 $\times$15$\times$1.5 mm$^3$, with 1500 connections for each transceiver IC feeding an 8$\times$4 patch antenna arras to support linear dual-polarization with 20 dB isolation and dual-frequency antennas. This work includes the validation of the design and packaging concept suitable for low-cost 5G mm-wave CPE and base-station applications. Qualcomm has been pioneering 5G UE modules with ICs in pair with 1$\times$4 dipole, 1$\times$4 patch, 2$\times$2 patch, and 2$\times$4 patch antenna arrays \cite{2018Dunworth}. The second generation of their antenna SiP module covered dual-band (28 and 39 GHz) operation with the size of 19.1$\times$4.9$\times$1.78 mm$^3$. The antenna module includes power-module IC and transceiver IC, which are flip-chip assembled on the other side of the antenna patterns with copper pillars.

    \begin{figure}[b]
		\centering
		\includegraphics[width=0.48\textwidth]{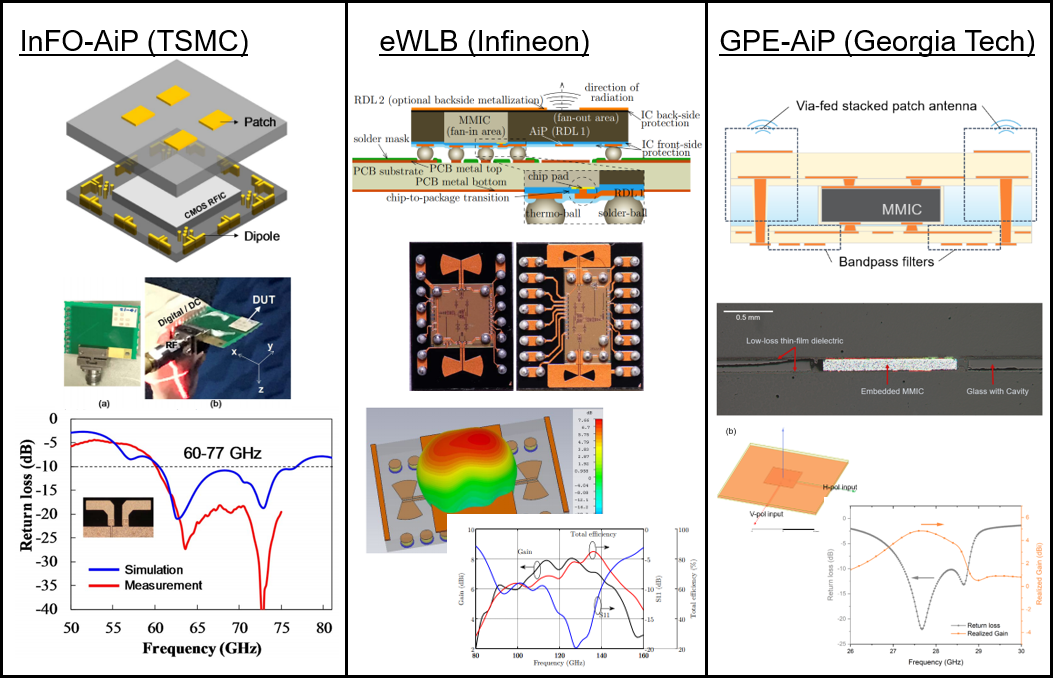}
    	\caption{Demonstrated mm-wave modules with IC-embedded AiP: molding-compound-based InFO-AiP from TSMC \cite{2019Tsai_iedmi}, eWLB from Infineon \cite{2018Ahmed}, and glass-based \cite{2020Watanabe_ectc} fan-out or chip-embedded packaging structures.}
		\centering
		\label{fig:fanout}
    \end{figure}

As one of the earlier demonstrations for AiP, a 60-GHz antenna integration with transceiver chips in a multi-layer LTCC were achieved by NEC Corporation, IBM \cite{2011Natarajan,2011Kam}, Samsung \cite{2011Hong}, Intel \cite{2013Kohen}, and a few more companies. Recently, Peraso Technologies Inc. in conjunction with Hitachi Metals Ltd. \cite{2019Rashidan} introduced novel corrugated soft or high-impedance surfaces and implemented them between phased-array antenna elements in LTCC procedure for 5G mm-wave communications. Their array preserves 14 dBi gain with 8 elements and 16.5 dBi gain with 16 elements over a 10-GHz bandwidth around 60 GHz. Glass-based AiP is also widely investigated especially in recent years \cite{2014Kamgaing,2018Watanabe,2020Xia,2020Zhang}, because of its potential of electrical properties, dimensional stability, and panel scalability toward 500$\times$500 mm$^2$. 

For the next-generation AiP, chip-embedded structures are promising for form-factor and thickness reduction with shorter interconnects, as shown in Fig. \ref{fig:fanout}. As discussed in Section \ref{subsec:trend}, Wojnowski and Bock from Infineon, for the first time, reported the fan-out wafer-level packaging (FOWLP) technology with their eWLB and presented antenna-integrated package in 2012 and 2015 \cite{2012Wojnowski,2015Bock}. Advanced Semiconductor Engineering (ASE) reported, in 2019, the FOWLP processes for stacked-patch antennas to enhance the bandwidth \cite{2019Hsieh}. TSMC has reported InFO-based antenna-integrated packages with beamforming capability of the antenna array system with 6 dBi gain in a 40-nm CMOS RFIC co-designed system \cite{2019Tsai_iedmi}. Tsai \textit{et al.} reported an antenna-integrated wafer-level package with a size of $10\times10\times0.5\; \text{mm}^3$, which shows an antenna-array gain of 14.7 dBi at the 60 GHz band \cite{2013Tsai}.  In addition to the wafer-level packaging, several R\&D teams from industry and academic research groups are investigating the potential of fan-out panel-level packaging (FOPLP) with organic substrates or glass substrates \cite{2019Ogura,2020Watanabe_ectc} for 5G and radar communications.  

\begin{figure}[b]
	\centering
	\includegraphics[width=0.35\textwidth]{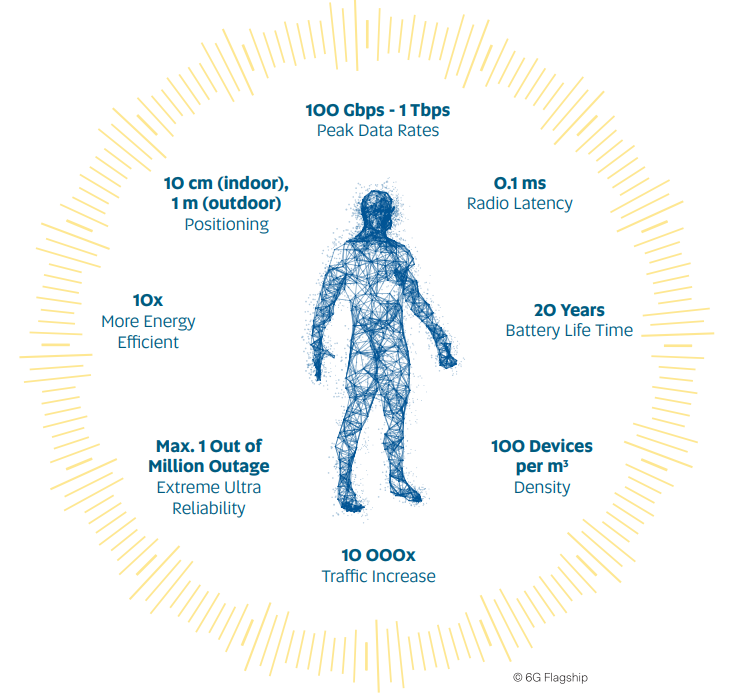}
	\caption{6G KPIs compared to 5G \cite{6GFlagship}.}
	\centering
	\label{fig:6G_specs_compared_to_5G}
\end{figure}

\section{5G and Beyond -- 6G}
\label{sec:6G}
6G is expected to utilize even higher frequencies and channel bandwidths than 5G, primarily by operating in the terahertz (THz) gap frequency range, resulting in massive data rates of 100 Gbps - 1 Tbps, compared to 10-20 Gbps of 5G \cite{First6G_WhitePaper}. The research in the domain of sub-THz to evaluate applications and use-cases is already underway in many focused research groups throughout the world \cite{6GFlagship, 6Genesis_AriPouttu}. 5G, as discussed in this paper, is known for its revolutionary flexibility. However, 6G will likely be known for using artificial intelligence (AI) by capitalizing on the backbone of flexibility offered by 5G. The other two use-cases are also expected to significantly advance with the advent of 6G: massive URLLC to provide true microsecond latency and mMTC to connect hundred of billions of devices. Some of the predicted 6G key performance indicators (KPI) compared to 5G are shown in Fig. \ref{fig:6G_specs_compared_to_5G}. A capability factor of 10-100 is expected for 6G as compared to 5G. The key technologies to enable 6G are AI, advanced RF, optical and network technologies \cite{R_W_5G_evol_path_2_6G, potential_key_tech_6G_mobile}. As a result, we can expect dominance of pervasive AI and machine learning to create smart, self-configurable mesh networks to support scalability with minimal human interaction. With THz radios, smart sensors and IoT driven by requirements such as HD artificial /virtual reality (AR/VR), extended reality (XR), local processing (wireless AI) and others, it would be imperative to identify potential new technologies to realize it, support evolution of use-cases and identify new ones to provide a ubiquitous, ultra-low latency, high-fidelity network regardless of the nature of environment \cite{Vision6G_wireless, 6G_Challenge_and_opportunity_survey}. 

\begin{figure}[t]
	\centering
	\includegraphics[width=0.48\textwidth]{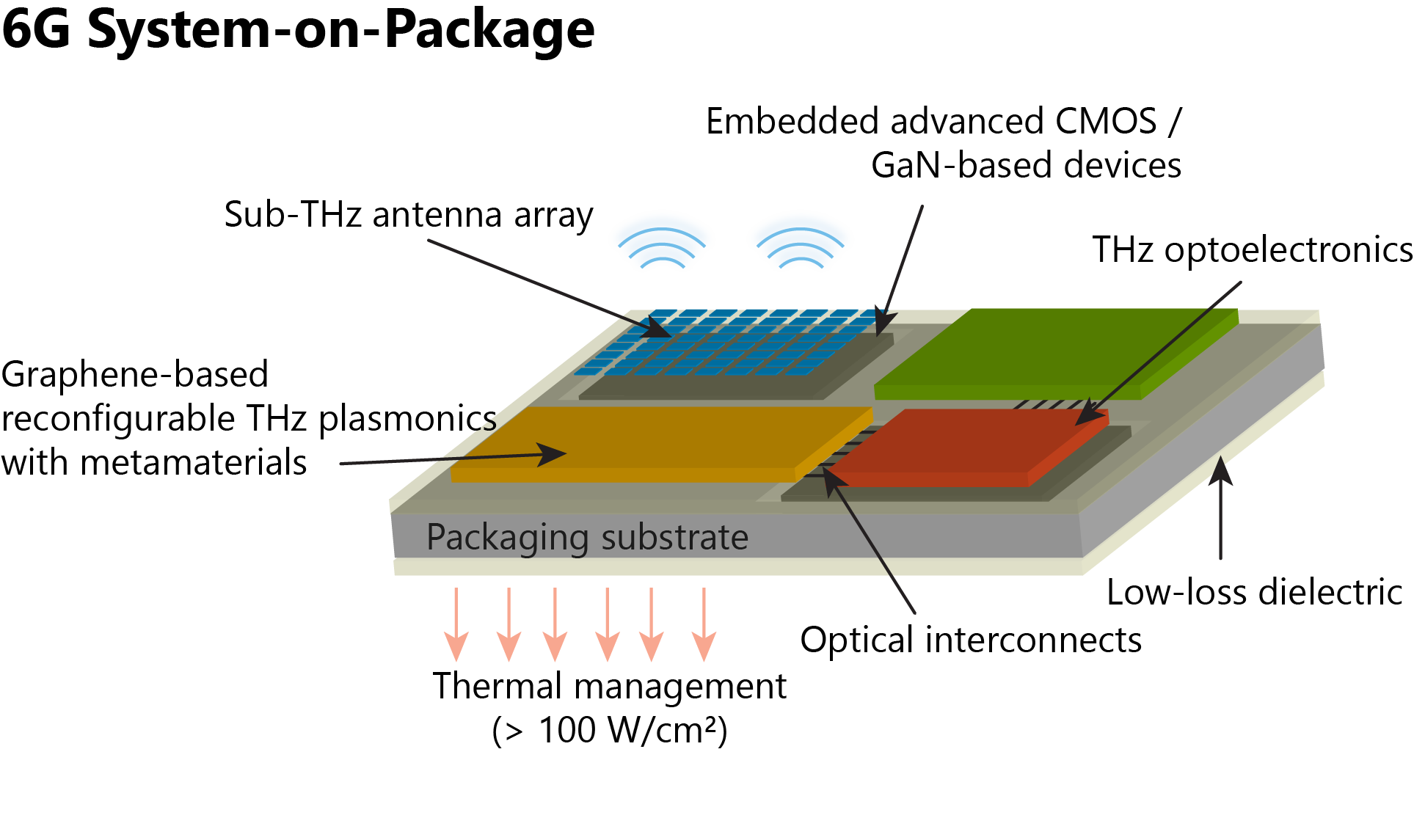}
	\caption{Conceptual diagram of heterogeneously-integrated quasi-optical THz package for 6G communications.}
	\centering
	\label{fig:6g_pkg}
\end{figure}

Many of the traditional approaches used for RF packages are not applicable to sub-THz and THz radios. Sub-THz or 6G communications focus on heterogeneous package integration, as illustrated in Fig. \ref{fig:6g_pkg} by incrementally advancing the system components such as precision antenna arrays, low-loss interconnects and waveguides and active devices. As frequency increases into the THz region, we will not be able to employ the techniques used for low frequencies because of multiple challenges associated with signal loss, dimensions, and materials. The current interconnection techniques such as wire bonding or solder-based bumps are too bulky; therefore, excitation of multimode, radiation, and reflection would adversely affect the electrical performance. As the skin depth is approximately 120 nm at 300 GHz, higher frequency packaging will require shorter interconnects and smooth surface to mitigate conductor loss. Radiation loss must be minimized as well with good impedance matching and high-precision manufacturing with tolerance below 1 $\mu$m. Dielectric loss is more predominant in sub-THz or THz bands, which translates to the need for accurate characterization of potential materials at those high frequencies. Another key metric in THz applications is antenna implementation. While AiP approach is viable, antennas or radiators on ICs will play a significant role. The integrated antenna may include graphene-based radiation sources and detectors, THz antenna arrays (e.g., 1024 elements in 1 mm$^2$), lenses, and intelligent metasurfaces with space-time-frequency coding and low-loss interconnects.  
Fig. \ref{fig:6g_pkg} shows a conceptual diagram of heterogeneously-integrated quasi-optical THz 6G package for a device with an antenna, where sub-THz or THz waves will be radiated into the air from reconfigurable AiP or the on-chip antenna \cite{2017Song}. Technical challenges such as heat management and EMI shielding must be also addressed for such high-frequency systems.  

\section{Summary}
\label{sec:conclusion}
Millimeter wave and THz-enabled sub-systems will dominate future communication networks for broadband wireless mobile connectivity, latency-sensitive applications such as vehicle-to-vehicle and vehicle-to-network,  and massive connectivity needs in IoT. They will also create mm-wave imaging, sensing, and other applications that simultaneously demand high bandwidth, reliability and zero perceived latency. The associated product segments are classified into base-station, customer premise equipment and user equipment. Despite the wide range of power ($<$ 1 W to 30 W), communication range (few meters to 100s of meters) and bandwidths requirements,  underlying technologies have several similarities in the package architectures, materials and processes.  The key challenges with mm wave communications, such as large path losses, interconnect losses, power-hungry devices, and other technology limits have been systematically addressed with advances in III-V devices, IC-package co-design, innovative beamforming architectures, and test methodologies, making 5G systems a reality today. The fundamental technologies are systematically classified and reviewed in this paper.

Unlike with 4G, antenna arrays need to be integrated in the 3D package with smaller sizes making antenna integration as the mainstream 5G package architecture. Multilayer organic (MLO) packages with transceiver dies flip-chip-attached to the backside of the package and antenna arrays on the top side, known as AiP phased-array system is widely adapted in today’s UE and base station products. Ultra-low transmission losses, below 0.1 dB/mm, fine lines and spaces with 1--5\% process variations to meet stringent in-band loss and out-of-band rejection specifications has been the key packaging focus for 5G modules. Various passive elements such as filters, power dividers and other functions are embedded into the packages as distributed components. Organic packages, however, are limited in line-width control due to dimensional instability, requiring more layers with low process precision and high via-transition losses, resulting in thicker packages. This has been paving way to inorganic substrates such as glass as an alternative to meet the density and impedance continuities across the packages.

THz or 6G communications are the emerging as the next frontier with 100--1000X increase in network speed compared to 4G and LTE, leading to new applications based on virtual reality, and intense high-resolution video communications and connectivity of a much larger number of devices, such as sensors for IoT. Compact and high-output power sources and high-responsivity low-noise detectors that can operate at THz band are required to surmount the high path-loss at these frequencies. Key innovations such as integration of graphene-based radiation sources and detectors, THz antenna arrays, lenses and intelligent meta-surfaces with space-time-frequency coding and low-loss interconnects will create the next set of R\&D challenges for academia and industry.

% if have a single appendix:
%\appendix[Proof of the Zonklar Equations]
% or
%\appendix  % for no appendix heading
% do not use \section anymore after \appendix, only \section*
% is possibly needed

% use appendices with more than one appendix
% then use \section to start each appendix
% you must declare a \section before using any
% \subsection or using \label (\appendices by itself
% starts a section numbered zero.)
%

% you can choose not to have a title for an appendix
% if you want by leaving the argument blank

% use section* for acknowledgment
%\section*{Acknowledgment}
%\label{sec:acknowledg}

% Can use something like this to put references on a page
% by themselves when using endfloat and the captionsoff option.
\ifCLASSOPTIONcaptionsoff
  \newpage
\fi

% trigger a \newpage just before the given reference
% number - used to balance the columns on the last page
% adjust value as needed - may need to be readjusted if
% the document is modified later
%\IEEEtriggeratref{8}
% The "triggered" command can be changed if desired:
%\IEEEtriggercmd{\enlargethispage{-5in}}

% references section

% can use a bibliography generated by BibTeX as a .bbl file
% BibTeX documentation can be easily obtained at:
% http://mirror.ctan.org/biblio/bibtex/contrib/doc/
% The IEEEtran BibTeX style support page is at:
% http://www.michaelshell.org/tex/ieeetran/bibtex/
%\bibliographystyle{IEEEtran}
% argument is your BibTeX string definitions and bibliography database(s)
%\bibliography{IEEEabrv,../bib/paper}
%
% <OR> manually copy in the resultant .bbl file
% set second argument of \begin to the number of references
% (used to reserve space for the reference number labels box)
%\begin{thebibliography}{1}

%\bibitem{IEEEhowto:kopka}
%H.~Kopka and P.~W. Daly, \emph{A Guide to \LaTeX}, 3rd~ed.\hskip 1em plus
%  0.5em minus 0.4em\relax Harlow, England: Addison-Wesley, 1999.

%\end{thebibliography}

\bibliographystyle{IEEEtran}

\bibliography{IEEEreference}

% biography section
% 
% If you have an EPS/PDF photo (graphicx package needed) extra braces are
% needed around the contents of the optional argument to biography to prevent
% the LaTeX parser from getting confused when it sees the complicated
% \includegraphics command within an optional argument. (You could create
% your own custom macro containing the \includegraphics command to make things
% simpler here.)
%\begin{IEEEbiography}[{\includegraphics[width=1in,height=1.25in,clip,keepaspectratio]{mshell}}]{Michael Shell}
% or if you just want to reserve a space for a photo:

% that's all folks
\end{document}